\documentclass[onecolumn,letterpaper]{ibmresearch}

\usepackage{amsmath,amssymb,amsfonts,amsopn}
\usepackage{booktabs}
\usepackage{braket}
\usepackage{bm}
\usepackage[british]{babel}

\usepackage{csquotes}
\usepackage{float}
\usepackage{transparent}
\usepackage[section]{placeins}

\usepackage{caption}
\usepackage{graphicx}
\graphicspath{ {./images/}{./svg-inkscape/} }

\usepackage{subcaption}

\usepackage{multirow}
\usepackage{longtable}

\usepackage{wrapfig}

\usepackage{pdfpages}
\usepackage{xcolor}
\usepackage{siunitx}
\usepackage[toc,page]{appendix}

\usepackage{xcolor}

\usepackage[colorinlistoftodos, textsize=footnotesize]{todonotes}
\usepackage[markup=underlined, authormarkup=brackets]{changes}
\definechangesauthor[name={tte}, color=purple]{tte}

\usepackage[
  style=nature,
  maxcitenames=2,
  maxbibnames=10,
  backend=biber,
  autocite=superscript,
  ]{biblatex}

\author[1,2]{Doriela Grabocka}
\author[1]{Dimitrios Christofidellis}
\author[1,3]{Federico Zipoli}
\author[1]{Anna van Elst}
\author[1]{Patrick Ruch}
\author[1]{Antonio Foncubierta-Rodríguez}
\author[2]{Siddhartha Mishra}
\author[1]{Matteo Manica}

\affiliation{1}{IBM Research}
\affiliation{2}{ETH Z\"urich}
\affiliation{3}{NCCR Catalysis, Switzerland}

\correspondingauthor{Doriela Grabocka}{doriela.grabocka@ibm.com}

\title{AI for Parameter Selection in Sample-based Quantum Diagonalization}

\date{\today}

\begin{document}
\justifying
\begin{abstract}
Sample-based Quantum Diagonalization (SQD) methods provide a way to compute molecular ground-state energies, but their accuracy and runtime depend sensitively on configuration choices such as the subspace dimension, propagation timestep, and the allocation of a fixed sampling budget across circuits, for which no formal selection method exists. In this work, we explore whether oracle models can predict the quantum runtime and approximation error from the Hamiltonian and the candidate configuration. Using synthetic Hamiltonians with controllable spectral properties, we show that machine-learned models can map a Hamiltonian directly to the best-ranked configuration within a predefined grid for the SKQD algorithm, validated against classical Krylov propagation, achieving strong rank correlations between true and predicted error and runtime, Spearman $\rho = 0.969$ and $\rho=0.905$, respectively. Extending the approach to 50 molecular Hamiltonians (20–50 qubits), with reference ground-state energies from classical calculations, we train regressors that predict the simulation error and runtime for a given molecule–configuration pair ahead of execution, and classifiers to distinguish, from $600$ SqDRIFT quantum-hardware runs, whether a configuration achieves a ground-state energy within chemical accuracy ($1.6\times10^{-3}$ Ha $\approx 1$\,kcal/mol). For the molecular Hamiltonians, the models additionally provide an \emph{a priori} feasibility and resource estimate, predicting whether a molecule can reach chemical accuracy on current hardware and at what runtime, thus correctly classifying held-out molecules at the molecule level, achieving balanced accuracy $61.8\%$ in-distribution and $100\%$ for five larger held-out molecules.
\end{abstract}

\maketitle

\section{Introduction}

\begin{figure}[htbp]
    \centering
    {\def\svgwidth{\linewidth}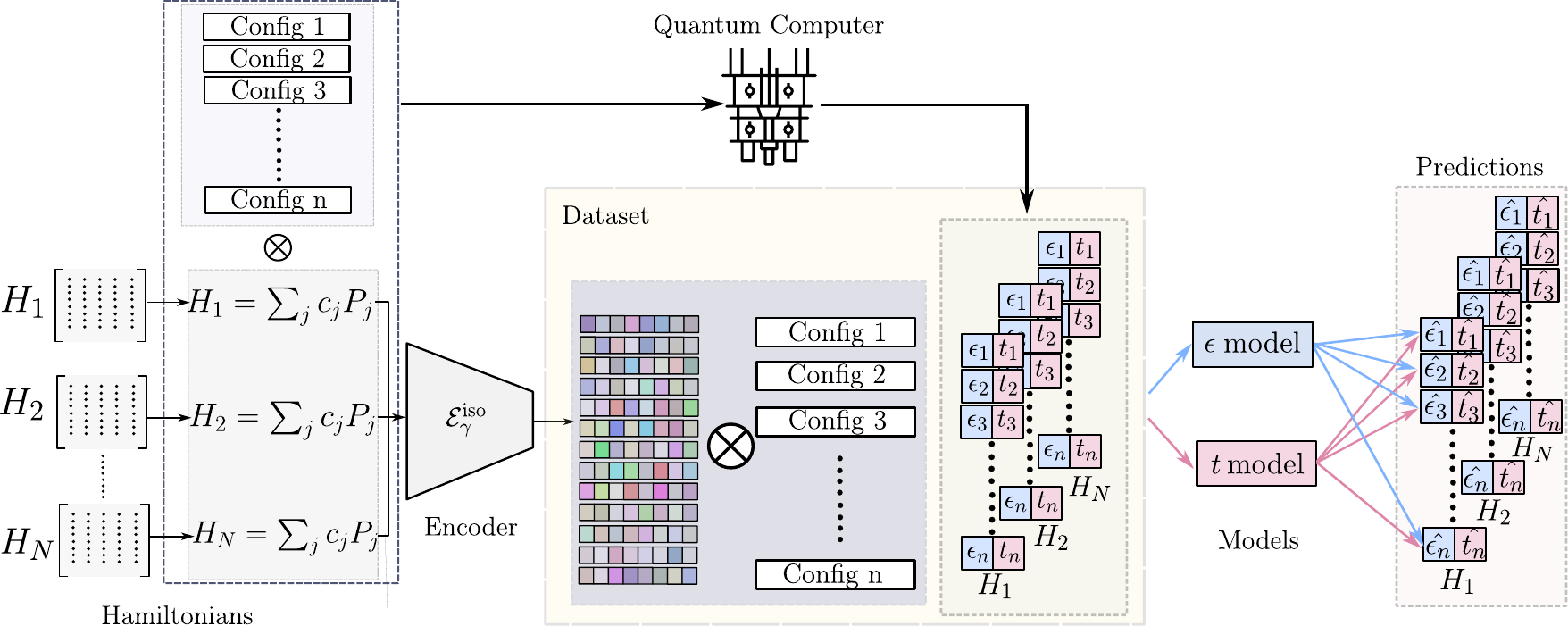}
    \caption{\textbf{The main pipeline.} The dataset consists of Hamiltonians that are decomposed into Pauli terms. The Pauli terms are used as input for the encoder, that creates 768-dimensional embeddings for each Hamiltonian. The embeddings are paired with all the configurations. The Hamiltonians and the configurations are executed on the quantum computer, yielding \texttt{runtime} and ground state energy approximation \texttt{error}. The embeddings of the Hamiltonians, combined with the configurations and the results from the quantum runs, create the dataset for training the \texttt{error} and \texttt{runtime} models to predict the ground state energy error and the quantum processing unit runtime.}
    \label{fig:main_figure}
\end{figure}
 
Ground state energies govern the behavior and observable properties of many chemical and physical systems. The ground state energy of a system corresponds to the lowest eigenvalue of its Hamiltonian, the Hermitian matrix that describes the system. Classical methods struggle to diagonalize very large matrices because the dimension of the Hilbert space grows exponentially with the system size, rendering exact diagonalization intractable. This motivates the development of Quantum Diagonalization Algorithms (QDAs), a promising approach for estimating ground-state energies on quantum computers~\autocite{lee2023evidence,reinholdt2025critical}. Algorithms such as Variational Quantum Eigensolvers (VQE)~\autocite{peruzzo_variational_2014} and Sample-based Quantum Diagonalization (SQD) methods~\autocite{yu_quantum-centric_2025, yoshioka_krylov_2025} are hybrid quantum-classical approaches for finding the ground state energy of a physical system. VQE methods rely on iterative variational optimization, which can be challenging on current Noisy Intermediate-Scale Quantum (NISQ)~\autocite{preskill2018nisq} devices due to both noise and optimization-related difficulties. In contrast, Krylov-based approaches such as Sample-based Krylov Quantum Diagonalization (SKQD)~\autocite{yu_quantum-centric_2025} and SqDRIFT~\autocite{piccinelli_quantum_2026} avoid variational optimization by constructing and diagonalizing a low-dimensional subspace.

SQD methods such as SKQD and SqDRIFT rest on the idea that only a subset of the determinants spanning the Hilbert space is relevant for computing the ground state energy of the system. Therefore, they sample quantum circuits to construct a lower-dimensional subspace and diagonalize the projected Hamiltonian within it, with convergence to the true ground state energy guaranteed provided that the ground state is sparse and there is significant overlap between the true ground state and the state sampled from the quantum circuit. The quantum device identifies an appropriate subspace, after which the diagonalization of the projected Hamiltonian proceeds entirely classically. The quality of the approximation and the efficiency of the quantum runs depend on the Hamiltonian and on the set of parameters with which the algorithm runs and its circuits are built. At present, neither a one-size-fits-all solution nor a standard method for choosing the appropriate parameters for a given Hamiltonian exists. In most cases, parameter selection is a trial-and-error procedure that costs time and resources. The SKQD and SqDRIFT approaches are described in detail in \autoref{sec:sup_skqd} and \autoref{sec:sup_sqdrift}.

In this work, we explore whether machine learning models can learn the relationship between Hamiltonian structure and algorithmic parameters on the one hand, and the resulting \texttt{runtime} and approximation \texttt{error} on the other. In this way, the preferred configuration for a specific problem can be estimated a priori, without requiring extensive quantum experiments solely for parameter optimization. Such models provide a foundation for data-driven parameter selection and resource estimation of quantum runs. We conduct the experiments in two stages: first, we apply the methods to a dataset of synthetic Hamiltonians of sizes $N=2^3$ (3 qubits) to $N=2^{10}$ (10 qubits); second, we validate them further on a dataset of molecular Hamiltonians representing molecules with 10 to 25 spatial orbitals (20 to 50 qubits), corresponding to Hamiltonians of sizes $N=2^{20}$ to $N=2^{50}$. In the first stage we simulate SKQD circuits classically, whereas in the second we test the models against actual hardware runs of SqDRIFT. 

An attractive property of Hamiltonians is that they can be expressed as a sum of Pauli terms, $H = \sum_{i=1}^{N} c_iP_i$, where $c_i \in \mathbb{R}$ and $P_i$ is the tensor product of the Pauli operators $X, Y, Z, I$. The proposed pipeline builds on this representation, as shown in~\autoref{fig:main_figure}: the Pauli terms are encoded and combined with the configurations to train a model that predicts the \texttt{error} and \texttt{runtime} of a quantum-hardware execution.

\paragraph{Related work.}
Sample-based quantum diagonalization originates from quantum-selected configuration interaction (QSCI)~\autocite{kanno2023qsci} and its quantum-centric-supercomputing realization with self-consistent configuration
recovery~\autocite{robledomoreno2025sqd}, and belongs to the broader family of quantum subspace methods~\autocite{motta2024subspace}. It has since been extended
to excited states~\autocite{barison2025extsqd} and embedded
fragments~\autocite{shajan2025dmetsqd}. Quantum Krylov approaches build the subspace from time-evolved states~\autocite{parrish2019qfd,stair2020mrqk,
klymko2022rte,cortes2022krylov}, with convergence analyses that make the dependence on the Krylov dimension and the timestep
explicit~\autocite{epperly2022theory,kirby2023lanczos}. SKQD~\autocite{yu_quantum-centric_2025,
yoshioka_krylov_2025} and SqDRIFT~\autocite{piccinelli_quantum_2026}, which replaces Trotterized propagators with qDRIFT randomized compilation~\autocite{campbell_random_2019}, bring these ideas into sample-based pipelines, alongside closely related time-evolution-based QSCI variants~\autocite{sugisaki2025hsbqsci,mikkelsen2025teqsci}. Critical assessments show that the sampling budget and the compactness of the sampled subspace strongly limit the attainable accuracy~\autocite{reinholdt2025critical}, and
hardware studies report a marked sensitivity to execution
settings~\autocite{bhuiyan2026robustness,raisuddin2025promise}, which motivates principled configuration choices.
 
Machine learning has so far entered SQD mainly at the level of configuration recovery~\autocite{zeni2026csqd,patra2025pigen}. For variational algorithms, learned models have been used to initialize or transfer circuit parameters across problem
instances~\autocite{verdon2019learning,cervera2021metavqe,sauvage2021flip,jain2022gnn,shaydulin2023transfer}. Closer in spirit to our work, supervised models have been trained to select compilation options and target devices~\autocite{quetschlich2023options,quetschlich2024mqtpredictor} and to predict circuit execution times~\autocite{ma2026exectime,xing2025qputime} and fidelities~\autocite{vadali2024fidelity}, and machine learning has been shown to map Hamiltonian parameters to ground-state properties with provable guarantees~\autocite{huang2022provably}. More broadly, our approach is an instance of empirical performance modeling for algorithm selection and configuration~\autocite{rice1976algorithm,leytonbrown2009empirical,xu2008satzilla,hutter2014runtime}, with precedents in the selection of sparse linear solvers from matrix features~\autocite{bhowmick2011solver,
jessup2016lighthouse} and in predicting the cost and outcome of classical quantum-chemistry calculations~\autocite{heinen2020cost,duan2019failure}. To our
knowledge, no prior work predicts the runtime and accuracy of SQD-family algorithms from the Hamiltonian and the algorithm configuration.

\paragraph{Contributions.}
SQD-based approximations of the ground state and its energy depend on the choice of circuit parameters, which in turn depends on the nature of the Hamiltonian. An ``oracle'' that indicates the preferred circuit configuration for the most accurate and efficient solution is therefore desirable. In this work, we demonstrate the benefits of learning to predict the outputs of SQD circuits across configurations. Specifically, our contributions are as follows: (i) we assemble datasets of synthetic and molecular Hamiltonians by simulating quantum runs classically and executing runs on real quantum hardware; (ii) we train an encoder that encodes Hamiltonians in a latent space based on their spectral properties; (iii) we train regressors that predict the \texttt{runtime} and ground state energy approximation \texttt{error} for a Hamiltonian $H$ and a set of parameters of an SQD algorithm; and (iv) we demonstrate model performance on two SQD algorithms, SKQD and SqDRIFT.

\section{Results}\label{sec:results}
\subsection{Datasets}\label{sec:r_datasets}
For training and validating the models for parameter selection, we consider two datasets. The first is a dataset of synthetically generated Hamiltonians, used for the SKQD circuits; the second is a dataset of molecular Hamiltonians used for the SqDRIFT circuits. 

The synthetic dataset also serves to train an encoder that learns to represent Hamiltonians in a latent space. The Hamiltonians used to train the encoder range in size from $N=2^{3}$ to $N=2^{72}$. A subset of the test set used to evaluate the encoder further trains the downstream models. These Hamiltonians of sizes $N=2^{3}$ to $N=2^{10}$ serve for training and evaluating the machine learning models that predict the \texttt{runtime} and \texttt{error}, given the Hamiltonian representation, the Krylov dimension $m$ and timestep $\Delta t$. We use the \texttt{runtime} and the state-propagation \texttt{error} of Krylov-Arnoldi~\autocite{saad1992expm,hochbruck1997krylov} real-time evolutions as a classical surrogate: the state-propagation error is a classical matrix-exponential error, related to but not identical with SKQD accuracy, and the runtime is classical wall-clock time rather than QPU runtime.

Second, a dataset of molecular Hamiltonians derived from 50 molecules is used to generate SqDRIFT circuits with different configurations for execution on quantum hardware ~\autocite{molecules_for_sqd_benchmarking}.
These molecules belong to 13 chemical classes and are represented by circuits ranging from 20 to 50 qubits (sizes from $N=2^{20}$ to $N=2^{50}$), with average two-qubit gate depths between 296 and 858 for the runs on the Quantum Processing Units (QPUs;~\autoref{sec:sup_chemistry_ham}). We compute the approximate ground state energies of these 50 molecules with the Truncated Davidson method~\autocite{cotton_truncated_2022,DAVIDSON197587}.

\subsection{Hamiltonian representation}\label{sec:r_encoder}

\begin{figure}[htbp]
    \centering
    \includegraphics[width=\linewidth]{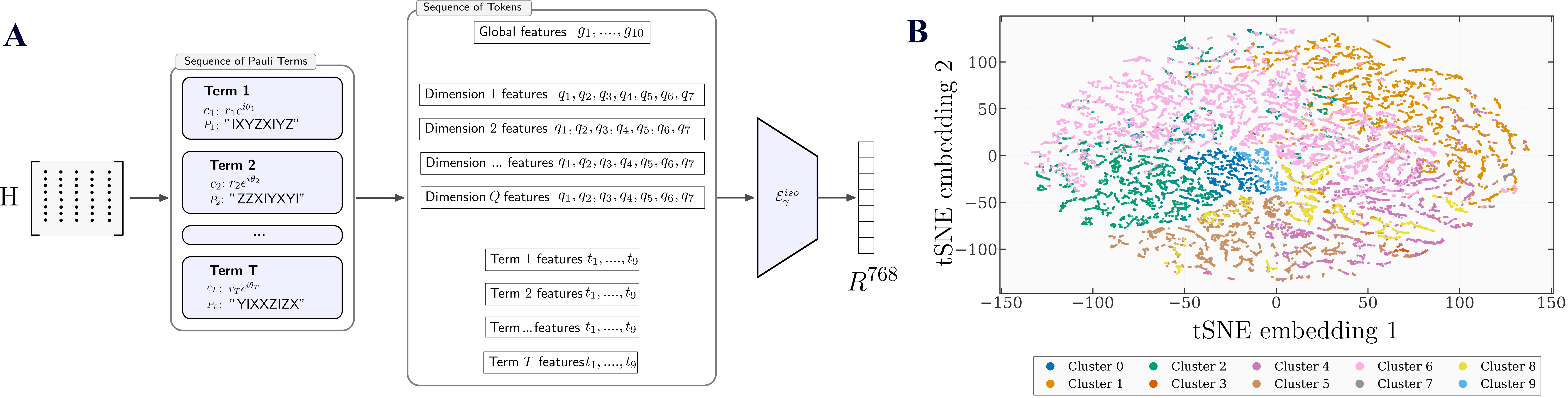}
    \caption{\textbf{The encoder model.} Part A shows the pre-processing of a Hamiltonian $H$. $H$ is expressed as a sum of Pauli terms, from which features are extracted. These Pauli terms' features are then concatenated with the global system features and qubit features, which are tokenized and input into the encoder model, which in turn generates the embeddings. For more details on the encoder architecture, refer to \autoref{sec:sup_encoder_architecture}. Part B shows how the embeddings of Hamiltonians with different spectra are scattered in a low-dimensional space using t-SNE. The spatial coherence of colour reflects how well the encoder groups spectrally similar matrices. Indices denote regions of a continuous spectral family rather than distinct physical classes; t-SNE axes are in arbitrary units (perplexity set to 20). Matrices sharing an eigenvalue spectrum are mapped to near-coincident points despite differing in structure (mean pairwise cosine distance 0.017, $n = 1268$ spectral classes over 44,514 matrices), while matrices with distinct spectra are well separated (0.681).}
    \label{fig:res_encoder_panel}
\end{figure}
   
  An encoder pre-trained on the synthetic-Hamiltonian dataset extracts features from the Hamiltonians for training the machine learning models. It takes as input the Pauli decomposition of a Hamiltonian and outputs a 768-dimensional embedding representing it. The Pauli terms, together with system-level and qubit features, are each linearly projected into a shared token sequence, marked by learned type embeddings, and used to train the encoder to represent Hamiltonians in latent space. \autoref{fig:res_encoder_panel} shows the preprocessing pipeline of the Hamiltonians (Panel A) and the t-SNE~\autocite{vanDerMaaten2008} low-dimensional representation of the 768-dimensional embeddings (Panel B). Panel B shows that the embeddings of Hamiltonians with similar spectra cluster together (same color), while those of Hamiltonians with different spectra lie further apart. For further details on the Hamiltonian representation and encoder training, see \autoref{sec:methodology_train_enc}.

\subsection{Performance on the synthetic Hamiltonians dataset}\label{sec:regression_on_synthetic_ham}
For dense synthetic Hamiltonians the time-evolution circuits would be prohibitively deep, so executing their SKQD circuits on real hardware is currently infeasible on NISQ devices. As a classical surrogate for this setting, we use the \texttt{runtime} and the state-propagation \texttt{error} of Krylov-Arnoldi~\autocite{saad1992expm,hochbruck1997krylov} real-time evolutions, where the state-propagation error is a classical matrix-exponential error that is related to, but not identical with, SKQD accuracy. This is a surrogate learning task: the goal is to show that the encoder together with the regressors can learn configuration-dependent performance on this proxy. We use the subset of the test set on which the encoder was evaluated to train two regressors that predict the state-propagation \texttt{error} and \texttt{runtime} for a Hamiltonian and a given $m$--$\Delta t$ configuration. To avoid data leakage, the train-test split is done at the level of the isospectral families. For each synthetic Hamiltonian we form a grid of 48 combinations, $m \in \{5,7,10,15,20,25\}$ and $\Delta t \in$
\{$0.01$, $0.02$, $0.05$, $0.1$, $0.2$, $0.5$, $1$, $2$\}. For details of how we compute the error, see \autoref{sup:skqd_simulations}.

\begin{figure}[htbp]
    \centering
    {\def\svgwidth{\linewidth}
\begingroup%
  \makeatletter%
  \providecommand\color[2][]{%
    \errmessage{(Inkscape) Color is used for the text in Inkscape, but the package 'color.sty' is not loaded}%
    \renewcommand\color[2][]{}%
  }%
  \providecommand\transparent[1]{%
    \errmessage{(Inkscape) Transparency is used (non-zero) for the text in Inkscape, but the package 'transparent.sty' is not loaded}%
    \renewcommand\transparent[1]{}%
  }%
  \providecommand\rotatebox[2]{#2}%
  \newcommand*\fsize{\dimexpr\f@size pt\relax}%
  \newcommand*\lineheight[1]{\fontsize{\fsize}{#1\fsize}\selectfont}%
  \ifx\svgwidth\undefined%
    \setlength{\unitlength}{585.85619829bp}%
    \ifx\svgscale\undefined%
      \relax%
    \else%
      \setlength{\unitlength}{\unitlength * \real{\svgscale}}%
    \fi%
  \else%
    \setlength{\unitlength}{\svgwidth}%
  \fi%
  \global\let\svgwidth\undefined%
  \global\let\svgscale\undefined%
  \makeatother%
  \begin{picture}(1,1.0802051)%
    \lineheight{1}%
    \setlength\tabcolsep{0pt}%
    \put(0,0){\includegraphics[width=\unitlength,page=1]{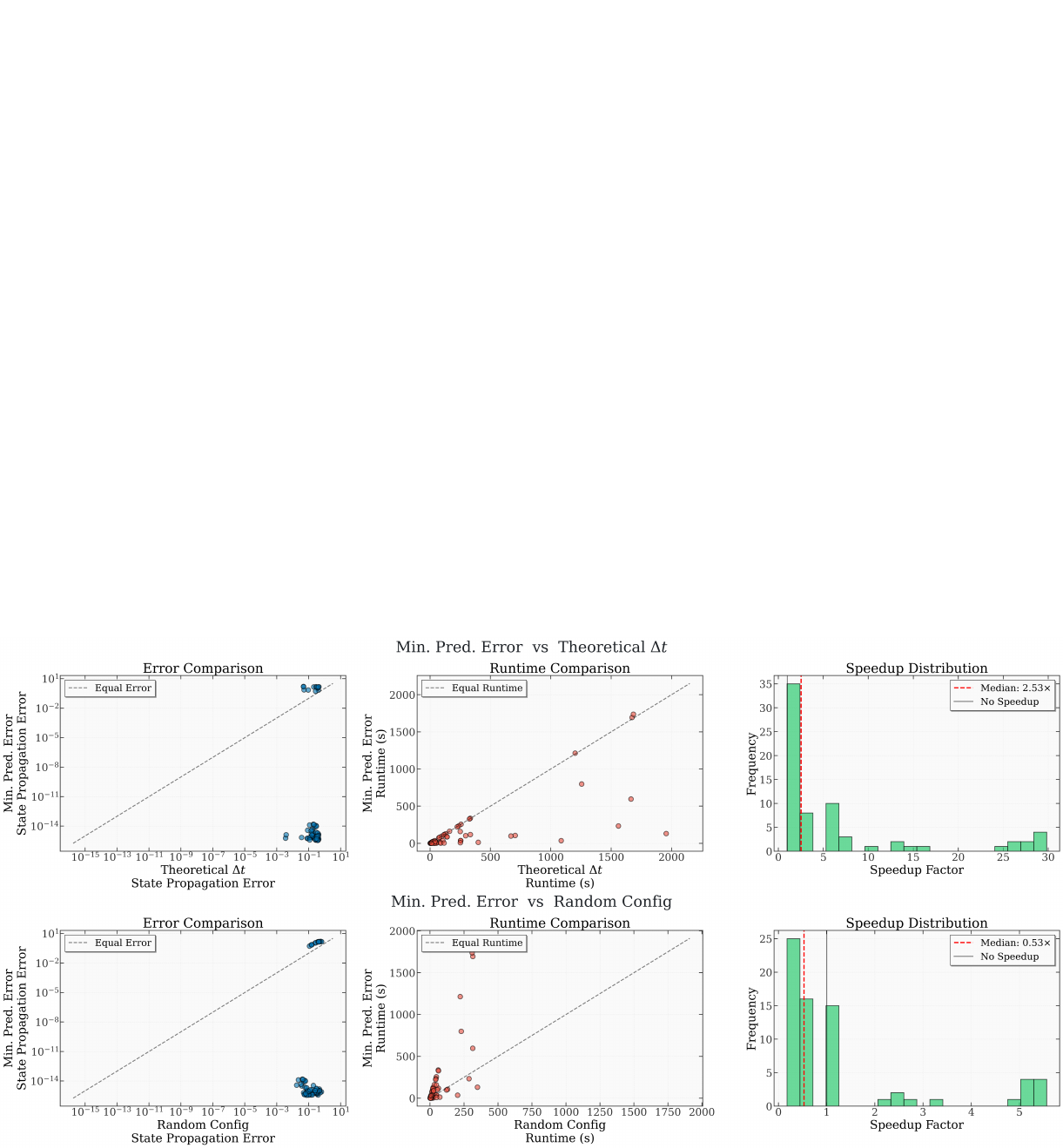}}%
    \put(0.00444874,0.46956727){\color[rgb]{0,0,0}\makebox(0,0)[lt]{\lineheight{0}\smash{\begin{tabular}[t]{l}\textbf{B}\end{tabular}}}}%
    \put(0,0){\includegraphics[width=\unitlength,page=2]{synthetic_results_panel2_svg-tex.pdf}}%
    \put(0.00208994,1.04530476){\color[rgb]{0,0,0}\makebox(0,0)[lt]{\lineheight{0}\smash{\begin{tabular}[t]{l}\textbf{A}\end{tabular}}}}%
    \put(0,0){\includegraphics[width=\unitlength,page=3]{synthetic_results_panel2_svg-tex.pdf}}%
  \end{picture}%
\endgroup%
}
    \caption{\textbf{Summary results on the synthetic-Hamiltonian dataset.} Panel A shows the test-set results for both the state-propagation \texttt{error} and \texttt{runtime} predictions. For training, we transform the targets: the runtime is log-transformed, while the error is mapped to a normal distribution using the \texttt{QuantileTransformer} from \emph{scikit-learn}~\autocite{pedregosa2011sklearn,sklearn_api}. We report the mean per-Hamiltonian Spearman correlation coefficient $\rho$ between true and predicted values, together with the mean per-Hamiltonian median absolute percentage error, on the original and transformed scales, to assess how well the transformed results transfer to the original scale. The vertical dashed line marks the dimensions present only in the test set. The third-column panels show scatter plots of predicted against true values. Panel B shows the results of the classical SKQD simulations using the \texttt{statevector} method of the \texttt{AerSimulator}. The plots compare the error and runtime of the model-suggested configuration (\emph{Minimum Predicted Error}) with those of a \emph{Theoretical $\Delta t$} configuration and a \emph{Random Configuration}.}
    \label{fig:res_synthetic_ham_results}
\end{figure}

\autoref{fig:res_synthetic_ham_results} shows the results for this set of experiments. Part A reports the per-dimension results of the regressors on the runtime and state-propagation error. During training, the models see only Hamiltonians up to dimension $N=2^7$ (7 qubits), whereas the test set contains Hamiltonians of all sizes from $N=2^3$ to $N=2^{10}$ (3 to 10 qubits). We train the models with a log-transformed runtime and a quantile-transformed~\autocite{sklearn_api} \texttt{error}, and therefore report the results on both the original and the transformed scales. We use two separate random forests of 500 trees, trained on 50 PCA features of the 768-dimensional embeddings of the synthetic Hamiltonians combined with the $m$--$\Delta t$ configuration. Because the ranking of configurations \emph{within} a given Hamiltonian is what matters for configuration selection, we report the mean per-Hamiltonian Spearman $\rho$ at each dimension, averaging over Hamiltonians the $\rho$ computed across the configurations of each one; the third-column scatter plots additionally show the pooled correlation over the whole test set. The per-Hamiltonian $\rho$ is high across dimensions for both \texttt{runtime} and \texttt{error} on both the transformed and the original scales, with a drop for \texttt{error} in the ``test-only'' regime. The median absolute percentage error of the \texttt{error} prediction is consistent across dimensions and scales, with matrix sizes seen during training yielding lower median absolute percentage errors than those not seen during training, though the median absolute percentage error remains below 50\% in all cases. By contrast, predicting the \texttt{runtime} for the larger dimensions unseen during training is more challenging than for the smaller ones, because the quantile-transformer-plus-random-forest predictions cannot exceed the range the model has seen during training. More important than the runtime error itself, however, is the ranking of the configurations, captured by the consistently high Spearman correlation coefficient $\rho$ across all dimensions. This indicates that the model ranks the configurations by runtime correctly.

The per-Hamiltonian values separate the two targets. For \texttt{runtime}, the mean per-Hamiltonian $\rho$ stays high at every size, ranging from $0.93$ to $0.98$, so the model ranks configurations by runtime correctly within each Hamiltonian, both in and out of distribution. For the \texttt{error}, the within-Hamiltonian ranking is strong at the sizes seen during training, mean per-Hamiltonian $\rho$ of $0.99$, $0.96$, $0.93$, and $0.89$ at $N=8$, $16$, $32$, and $64$, but degrades as the model extrapolates, falling to $0.78$ at $N=128$ and to $0.67$, $0.43$, and $0.61$ at the held-out sizes $N=256$, $512$, and $1024$. We therefore conclude that the model selects runtime-optimal configurations reliably across all sizes, and error-optimal configurations reliably at in-distribution sizes, with reduced reliability on the largest unseen Hamiltonians.

Part B reports the results of the classical SKQD simulations using the Qiskit \texttt{AerSimulator}~\autocite{qiskit2024, qiskitaddonsqd,robledomoreno2025sqd}. From the configurations the model suggests, we select the best, \emph{Minimum Predicted Error}, and compare it against a \emph{Random Configuration}, random combinations within the bounds of the $m$--$\Delta t$ grid; and a configuration combining the largest available $m=25$ with the theoretical $\Delta t$ of each Hamiltonian, computed as $\Delta t_{theor.} = \pi / \Delta E_{N-1}$ following~\autocite{yu_quantum-centric_2025}, where $\Delta E_{N-1}$ is the spectral range approximated via the Gershgorin bounds~\autocite{varga_gersgorin_2004}. In the first row of Part~B, the model-selected configuration achieves a lower classical state-propagation error, and a lower classical simulation \texttt{runtime}, than the theoretical-$\Delta t$ configuration, with a classical-simulation speedup of $2.53\times$ shown in the speedup-distribution plot. The random configurations run considerably faster than the model configurations, but the model-based configurations achieve a lower proxy error. Both comparisons indicate that the surrogate regressors predict the classical proxy error and runtime well and select configurations that minimize them; however, they are not a claim about SKQD performance on hardware. The configurations the model actually selects across dimensions are reported in
\autoref{tab:synthetic-selected-configs}.

Given these SKQD simulation results, we next evaluate the method on the more practical SqDRIFT circuits, which are executed on quantum hardware. The details follow in \autoref{sec:regression_on_real_ham}.

\subsection{Performance on the molecular Hamiltonians dataset}\label{sec:regression_on_real_ham}

\begin{figure}[htbp]
    \centering
    {\def\svgwidth{\linewidth}
\begingroup%
  \makeatletter%
  \providecommand\color[2][]{%
    \errmessage{(Inkscape) Color is used for the text in Inkscape, but the package 'color.sty' is not loaded}%
    \renewcommand\color[2][]{}%
  }%
  \providecommand\transparent[1]{%
    \errmessage{(Inkscape) Transparency is used (non-zero) for the text in Inkscape, but the package 'transparent.sty' is not loaded}%
    \renewcommand\transparent[1]{}%
  }%
  \providecommand\rotatebox[2]{#2}%
  \newcommand*\fsize{\dimexpr\f@size pt\relax}%
  \newcommand*\lineheight[1]{\fontsize{\fsize}{#1\fsize}\selectfont}%
  \ifx\svgwidth\undefined%
    \setlength{\unitlength}{572.6760648bp}%
    \ifx\svgscale\undefined%
      \relax%
    \else%
      \setlength{\unitlength}{\unitlength * \real{\svgscale}}%
    \fi%
  \else%
    \setlength{\unitlength}{\svgwidth}%
  \fi%
  \global\let\svgwidth\undefined%
  \global\let\svgscale\undefined%
  \makeatother%
  \begin{picture}(1,0.28850241)%
    \lineheight{1}%
    \setlength\tabcolsep{0pt}%
    \put(0,0){\includegraphics[width=\unitlength,page=1]{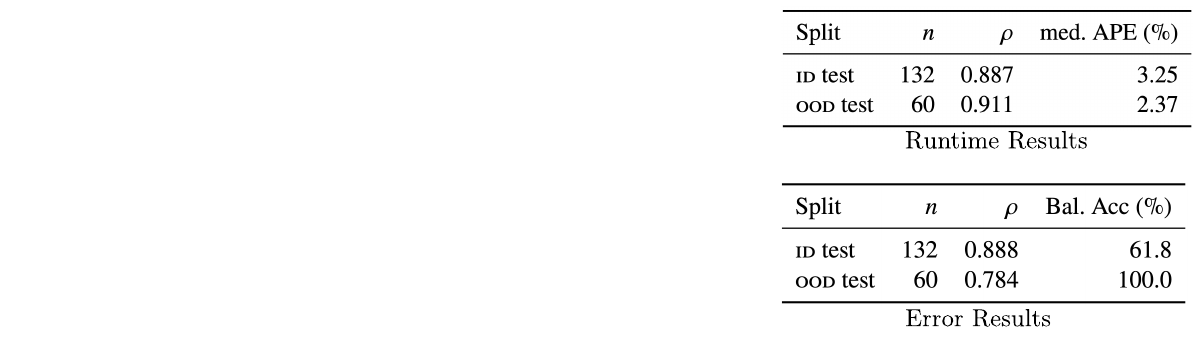}}%
    \put(0.62638638,0.25812761){\color[rgb]{0,0.03137255,0.21568627}\makebox(0,0)[lt]{\lineheight{0}\smash{\begin{tabular}[t]{l}\textbf{B}\end{tabular}}}}%
    \put(0,0){\includegraphics[width=\unitlength,page=2]{real_ham_results_test_only_svg-tex.pdf}}%
    \put(0.00048524,0.2634928){\color[rgb]{0,0.03137255,0.21568627}\makebox(0,0)[lt]{\lineheight{0}\smash{\begin{tabular}[t]{l}\textbf{A}\end{tabular}}}}%
  \end{picture}%
\endgroup%
}
    \caption{\textbf{Summary results on the molecular-Hamiltonian dataset.} Panel A shows the predicted versus true runtime and absolute error for the two test sets, in-distribution (ID) and out-of-distribution (OOD). The shaded area in the error plot highlights the samples whose true and predicted errors fall beyond chemical accuracy (CA${}=1.6 \times 10^{-3}$ Ha). Panel B reports, for the runtime, the Spearman correlation coefficient and the median absolute percentage error (med.\ APE) between true and predicted runtimes on the ID and OOD test sets; and, for the error, the Spearman correlation coefficient between true and predicted absolute errors together with the balanced accuracy for classifying whether a configuration yields a prediction below or above chemical accuracy.}
    \label{fig:res_real_ham_results}
\end{figure}

For this set of experiments, the SqDRIFT circuits for each of the 50 molecular Hamiltonians are run with 12 different configurations. We vary three parameters across these runs: \texttt{num\_groups} $\in \{15, 20\}$, \texttt{num\_rand} $\in \{100, 500, 1000\}$ and \texttt{m} $\in \{3,4\}$ with $\Delta t = 1$ (a detailed list of configurations, together with their effect on runtime and error, can be found in \autoref{sup:sqdrift_runs}). In total, we collect 600 data points over more than 95 QPU hours.

We split these 600 data points into three sets at the molecule level: the train set contains 34 molecules of 20 to 40 qubits, whose runs train the models; the in-distribution (ID) test set contains 11 molecules of 20 to 40 qubits; and the out-of-distribution (OOD) test set contains the runs for the molecules of 42 to 50 qubits, sizes the models never see during training. This split assesses the generalizability of the regressors to Hamiltonians of sizes beyond those seen during training.

We train two regressors, one for the \texttt{runtime} and one for the ground state energy approximation \texttt{error}. \autoref{fig:res_real_ham_results} shows the evaluation results. Panel A shows that the model predicts the runtime well on both test sets, achieving a Spearman correlation coefficient between true and predicted runtimes of $\rho=0.887$ on the ID test set and $\rho=0.911$ on the OOD test set. The \texttt{runtime} results in Panel B also show that the median absolute percentage error is low in both test sets, $3.25\%$ and $2.37\%$ for the ID and OOD test sets, respectively. The model that predicts the \texttt{runtime} is a random forest~\autocite{breiman2001rf} of 50 trees, using as features the full 768-dimensional embedding of the Hamiltonians combined with the configuration from the QPU runs.

Predicting the ground state energy approximation \texttt{error}, by contrast, proves to be more complex. Panel A of \autoref{fig:res_real_ham_results} shows the scatter plot of true and predicted errors, highlighting the region where errors go beyond the chemical accuracy at $1.6 \times 10^{-3}$ (grey zone). The \texttt{error} results in Panel B show that the Spearman correlation coefficient between the true and predicted errors is $\rho=0.888$ on the ID test set and $\rho=0.784$ on the OOD test set. This model is an Ordinary Least Squares model from \emph{scikit-learn}\autocite{pedregosa2011sklearn,sklearn_api}, using 8 PCA-reduced features of the 768-dimensional embedding together with the configuration from the QPU runs and four physical features: the number of qubits, the number of orbitals, the number of electrons, and $\log_{10} \text{CI}$ (the configuration-interaction, i.e.\ Hilbert-space, dimension). This again shows the importance of combining the Hamiltonian representation with the SqDRIFT circuit configuration for predicting the error.

In practice, the most useful question this classifier answers is one of \emph{a priori feasibility and resource estimation}: given a molecule and the available hardware, will any configuration reach chemical accuracy, and at what QPU cost?  Panel B therefore also reports the balanced accuracy obtained by reading the same regressor as a classifier, mapping its predicted error to a probability of exceeding the $1.6 \times 10^{-3}$ threshold. It reaches a balanced accuracy of $61.8\%$ on the ID (across configurations for 11 molecules) test set and $100\%$ on the OOD (across configurations for 5 molecules) test set. The two numbers describe different regimes rather than a contradiction: only 17 of the 132 ID test configurations lie beyond chemical accuracy, and there the model is strongly conservative, whereas on the OOD test set, where the classes are close to balanced (36 of 60 beyond the threshold), all 60 configurations of the five held-out molecules are classified correctly. For detailed results and ablations refer to ~\autoref{sup:ablations_molecular_hamiltonians}.

\subsection{Generalizing to out-of-grid circuit parameters}\label{sec:oog_parameters}

\begin{figure}[htbp]
    \centering
    {\def\svgwidth{\linewidth}
\begingroup%
  \makeatletter%
  \providecommand\color[2][]{%
    \errmessage{(Inkscape) Color is used for the text in Inkscape, but the package 'color.sty' is not loaded}%
    \renewcommand\color[2][]{}%
  }%
  \providecommand\transparent[1]{%
    \errmessage{(Inkscape) Transparency is used (non-zero) for the text in Inkscape, but the package 'transparent.sty' is not loaded}%
    \renewcommand\transparent[1]{}%
  }%
  \providecommand\rotatebox[2]{#2}%
  \newcommand*\fsize{\dimexpr\f@size pt\relax}%
  \newcommand*\lineheight[1]{\fontsize{\fsize}{#1\fsize}\selectfont}%
  \ifx\svgwidth\undefined%
    \setlength{\unitlength}{836.49564968bp}%
    \ifx\svgscale\undefined%
      \relax%
    \else%
      \setlength{\unitlength}{\unitlength * \real{\svgscale}}%
    \fi%
  \else%
    \setlength{\unitlength}{\svgwidth}%
  \fi%
  \global\let\svgwidth\undefined%
  \global\let\svgscale\undefined%
  \makeatother%
  \begin{picture}(1,0.39453567)%
    \lineheight{1}%
    \setlength\tabcolsep{0pt}%
    \put(0,0){\includegraphics[width=\unitlength,page=1]{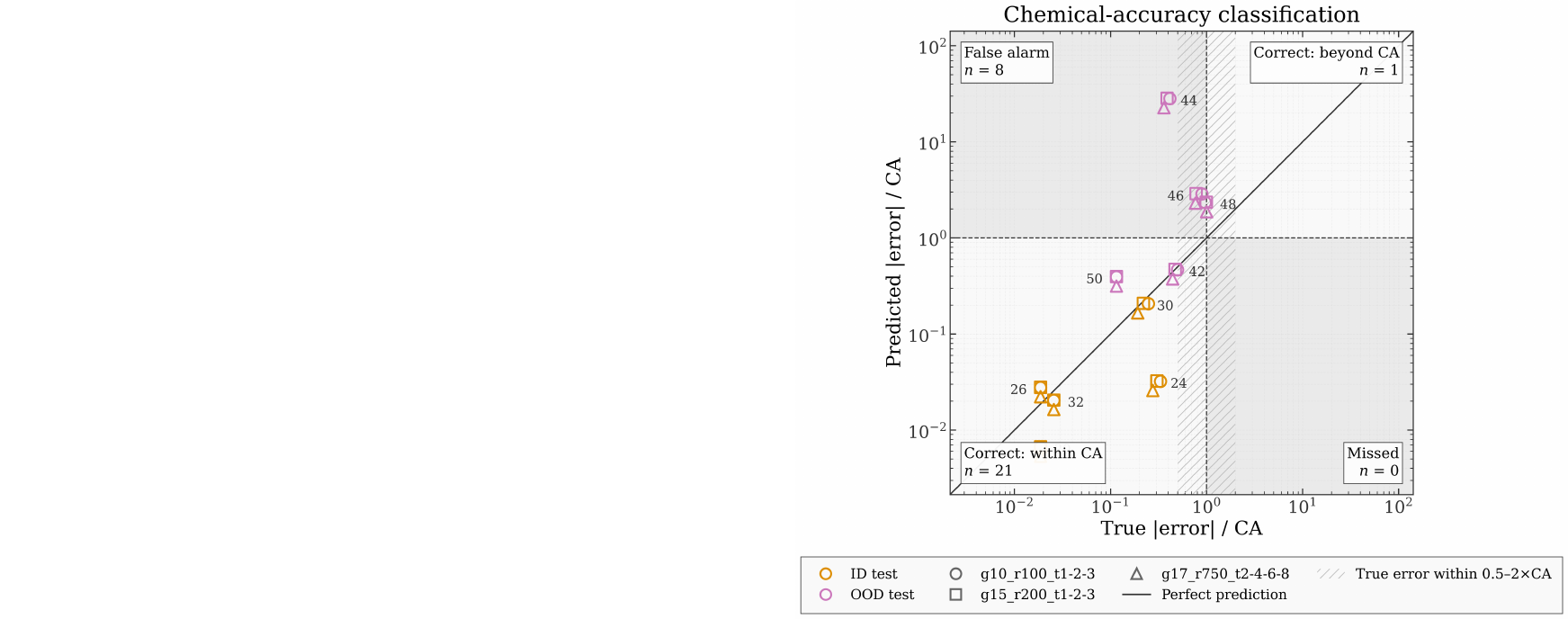}}%
    \put(0.55179165,0.36245855){\color[rgb]{0,0.04705882,0.29019608}\makebox(0,0)[lt]{\lineheight{0}\smash{\begin{tabular}[t]{l}\textbf{B}\end{tabular}}}}%
    \put(0,0){\includegraphics[width=\unitlength,page=2]{oog_results_svg-tex.pdf}}%
    \put(0.00744944,0.35845796){\color[rgb]{0,0.04705882,0.29019608}\makebox(0,0)[lt]{\lineheight{0}\smash{\begin{tabular}[t]{l}\textbf{A}\end{tabular}}}}%
    \put(0.01485847,0.09801952){\color[rgb]{0,0.04705882,0.29019608}\makebox(0,0)[lt]{\lineheight{0}\smash{\begin{tabular}[t]{l}\textbf{C}\end{tabular}}}}%
    \put(0,0){\includegraphics[width=\unitlength,page=3]{oog_results_svg-tex.pdf}}%
  \end{picture}%
\endgroup%
}
    \caption{\textbf{Out-of-grid circuit parameter generalization.} Part A shows the drift of the predicted runtime from the true runtime observed during the hardware execution. Part B plots the predicted error and the observed error, expressed as distance from chemical accuracy. The labels represent the number of qubits of the molecule. Part C is the table with the details of the out-of-grid circuit parameter configurations.}
    \label{fig:res_oog_parameters}
\end{figure}

In this section, we assess the generalization of the models from \autoref{sec:regression_on_real_ham} to configurations that are not seen during training. Five of the 11 molecules from the ID test set and the five molecules of the OOD test set are selected for 3 quantum hardware runs each on an IBM Heron r2 processor. Part C in \autoref{fig:res_oog_parameters} shows the exact configurations for the SqDRIFT circuits. The circuit parameters take the following values: \texttt{num\_groups} $\in \{10, 15, 17\}$, with 10 lying entirely outside the grid range $\{15, 20\}$, \texttt{num\_rand} $\in \{100, 200, 750\}$, with 200 and 750 being unseen values within the original range $\{100, 500, 1000\}$ and \texttt{m} $\in \{3,4\}$ with $\Delta t = 1$ for $m=3$ and $\Delta t = 2$ for $m=4$. The remaining parameters are kept the same as in \autoref{tab:sqdrift_configs}. 

Part A of \autoref{fig:res_oog_parameters} shows the drift of the predicted runtime from the hardware-observed runtime. The predicted runtime is usually between 0.89 and 1.28 times the true runtime. The most problematic configuration is the one that uses 10 groups, whose runtime is usually overpredicted. This, however, is an inherent limitation of random forests, which cannot predict values outside the range seen during training. It is worth noting here that \texttt{num\_groups}$=10$ is a value outside the range of the variable that was observed during training (\texttt{num\_groups}$=\{15, 20\}$). The predictive performance on the other two configurations is better, and the drift is smaller. These configurations are also unseen during training, but they fall within the ranges of values that the models saw during training.

Part B of \autoref{fig:res_oog_parameters} compares the predicted and measured errors, expressed as powers of 10 relative to chemical accuracy, since the errors span several orders of magnitude. The figure shows that the model correctly classifies all configurations for the ID molecules as giving errors within chemical accuracy. It correctly classifies the configurations two of five OOD molecules, including the largest one (50 qubits). For the other three molecules, it tends to overpredict the error. The model does not underpredict any errors, which would be worse than overprediction.

\section{Discussion}\label{sec:discussion}
In this work, we first show how to represent Hamiltonians with an encoder for downstream tasks, and then demonstrate the generalization abilities of regressors that predict quantum runtime and error for Hamiltonian-configuration pairs. The pre-trained encoder encodes both synthetic and molecular Hamiltonians, which are in turn paired with configurations of SKQD and SqDRIFT to assess the models' performance. The prediction performance on the dataset of synthetic Hamiltonians paired with configurations of the SKQD algorithm, where the model predicts both the proxy runtimes and errors reliably, is presented in ~\autoref{sec:results}. On the classical Krylov-Arnoldi proxy, the surrogate regressors select configurations that achieve a lower proxy error than the theoretical-$\Delta t$ and random baselines while reducing the classical simulation time by $2.5\times$. We also demonstrate the regressors' ability to predict the error and runtime of the configurations on quantum runs of SqDRIFT for 50 molecular Hamiltonians, achieving high ranking correlations of 0.888 and 0.784 for the ID (11 molecules) and OOD (5 larger molecules) test sets, respectively, with balanced accuracies for distinguishing configurations that achieve chemical accuracy in the energy error ranging $61.8\%$ to $100\%$ in the same sets. We therefore frame this part of the study as \emph{a priori feasibility and resource estimation}, in predicting whether a molecule can reach chemical accuracy on current hardware and at what QPU cost.

Despite the promising results, a main limitation is that the models are trained and evaluated mostly on configurations known at training time. Another limitation is that the models' predictions are tightly coupled to the devices where the quantum hardware experiments were run, so the transferability of the results to another device is not guaranteed. We make an initial effort to assess generalization to out-of-grid circuit parameters, with promising results; however, more inclusive and wider experiments are needed to strengthen this claim and address the limitations of the current models. A natural future direction is to explore the predictive performance of other models, that do not exhibit the constraints of the current ones on arbitrary configurations that do not appear during training, and to validate these predictions against quantum runs. Another future direction is to extend quantum-hardware validation to even larger molecules. The difficulty of such an experiment is that no classical reference for the ground-state is available for these molecules, which currently prevents us from expanding our dataset to larger Hamiltonians.

\section{Methods}\label{sec:methodology}
\subsection{Krylov--Arnoldi real-time evolution of an initial state}
\label{sec:methodology_krylov_arnoldi}

For the synthetic Hamiltonian dataset we generate the two regression targets,
\texttt{runtime} and state-propagation \texttt{error}, with a fixed-dimension
Krylov--Arnoldi propagator\autocite{saad1992expm,hochbruck1997krylov}, which
serve as classical proxies for the corresponding QPU quantities.

Given a Hermitian $H\in\mathbb{C}^{N\times N}$ and a normalized initial state
$\ket{\psi_0}$, the task is to approximate
\begin{equation}
    \ket{\psi(T)} = e^{-iHT}\ket{\psi_0}
    \label{eq:methods_time_evolution}
\end{equation}
at a fixed final time $T$. The evolution is carried out in $n=T/\Delta t$ uniform steps,
\begin{equation}
    \ket{\psi_{k+1}} \simeq e^{-i\Delta t H}\ket{\psi_k},
    \qquad k=0,\dots,n-1,
    \label{eq:methods_step}
\end{equation}
and each step is evaluated in a Krylov subspace of fixed dimension $m$. We
sweep $m\in\{5,7,10,15,20,25\}$ and
$\Delta t\in\{0.01,0.02,0.05,0.1,0.2,0.5,1,2\}$, giving $48$ combinations per matrix.

Because the propagator in \autoref{eq:methods_step} acts on the \emph{current} state, the relevant subspace at step $k$ is the order-$m$ polynomial Krylov space generated by $H$ and $\ket{\psi_k}$,
\begin{equation}
    \mathcal{K}_m(H,\psi_k)
      = \operatorname{span}\bigl\{\ket{\psi_k},\,H\ket{\psi_k},\,
        H^2\ket{\psi_k},\,\dots,\,H^{m-1}\ket{\psi_k}\bigr\}
    \label{eq:methods_krylov_subspace}
\end{equation}

The monomial basis in \autoref{eq:methods_krylov_subspace} is severely
ill-conditioned and is never formed. Instead, $m$ steps of the Arnoldi
iteration with modified Gram--Schmidt orthogonalization produce an orthonormal basis $V_m=[v_1,\dots,v_m]$ of $\mathcal{K}_m$, with
$v_1=\ket{\psi_k}/\beta_k$ and $\beta_k=\lVert\psi_k\rVert$, and return the Rayleigh--Ritz projection of the scaled generator onto that basis,
\begin{equation}
    \tilde H_m = V_m^{\dagger}\bigl(-i\Delta t H\bigr)V_m
    \in \mathbb{C}^{m\times m},
    \label{eq:methods_projection}
\end{equation}
which is upper Hessenberg and is assembled entry by entry from the orthogonalization coefficients rather than by a separate projection step. The iteration realises
$(-i\Delta t H)V_m = V_m\tilde H_m + \tilde h_{m+1,m}\,v_{m+1}e_m^{\top}$,
and the approximation consists in discarding the rank-one remainder. The propagated state is
\begin{equation}
    \ket{\psi_{k+1}} = \beta_k\,V_m\exp\!\bigl(\tilde H_m\bigr)e_1,
    \label{eq:methods_krylov_step}
\end{equation}
with $e_1$ the first canonical basis vector of $\mathbb{C}^m$. Equivalently,
$\ket{\psi_{k+1}}=q(-i\Delta t H)\ket{\psi_k}$ where $q$ is the degree-$(m-1)$
polynomial interpolating the exponential at the Ritz values of $\tilde H_m$;
fixing $m$ thus fixes the degree of the polynomial approximation to
$e^{-i\Delta t H}$. The scheme is thus a restarted Krylov propagator, with $V_m$ rebuilt at every step, rather than a single Krylov space spanning the whole evolution.

Denoting by $\ket{\psi^{(m,\Delta t)}(T)}$ the result of applying
\autoref{eq:methods_krylov_step} $n=T/\Delta t$ times, the state-propagation error is the $L_2$ distance to the exactly propagated state,
\begin{equation}
    \epsilon_{\mathrm{prop}}(m,\Delta t) =
    \bigl\lVert\,\ket{\psi^{(m,\Delta t)}(T)} - e^{-iHT}\ket{\psi_0}\,
    \bigr\rVert_2 ,
    \label{eq:methods_st_prop_error}
\end{equation}
where the reference $e^{-iHT}\ket{\psi_0}$ is obtained from a dense matrix exponential, which remains tractable at the dimensions considered here ($N\le 1024$). The initial state is a normalized complex Gaussian (Haar-random) vector, fixed by a common random seed so that all matrices of a given dimension and all $(m,\Delta t)$ pairs share the same $\ket{\psi_0}$.
The \texttt{runtime} target is the wall-clock time of the $n$-step loop alone, excluding construction of $H$ and of the reference state.

Alternatively, a quantum computer can sample the bitstrings that define the Krylov subspace and compute $\tilde{H}$. We therefore simulate the SKQD algorithm classically with multiple model-based configurations, random configurations, and configurations that use a theoretical $\Delta t$, computed as $\Delta t_{theor.} = \pi / \Delta E_{N-1}$ and combined with the largest $m$ from our grid, in order to compare their performance. The detailed steps of the SKQD algorithm are given in \autoref{sec:sup_skqd}, and details of the ranking of the simulations with these configurations are given in \autoref{sup:skqd_simulations}.

\subsection{Quantum hardware runs}\label{sec:methodology_quantum_hw_runs}

To create the SqDRIFT dataset, we use 50 molecular Hamiltonians ranging from 20 to 50 qubits and belonging to 13 chemical classes (details in \autoref{sec:sup_chemistry_ham}). We generate the corresponding SqDRIFT circuits for each of the 50 molecules by varying the number of Pauli terms subsampled from all $M$ Pauli terms of the Hamiltonian, \texttt{num\_groups}$=\{15,20\}$; the number of times the Pauli terms are subsampled, \texttt{num\_rand}$=\{100, 500, 1000\}$, which also equals the number of circuits; and the Krylov dimension, \texttt{m}=\{3,4\} with $\Delta t=1$. For each molecule, we perform 12 SqDRIFT runs on an IBM Quantum Heron r2 processor. Once the bitstrings from the quantum runs are collected, we filter them using configuration-recovery methods and project the Hamiltonian representing the molecule onto the subspace spanned by the recovered bitstrings. We then diagonalize the projected Hamiltonian to obtain its ground state energy. The difference between the true ground state energy of the molecule and the one obtained from the quantum runs determines the error $\epsilon$. The runtime $t$ is the runtime of the circuit in the QPU, excluding transpilation time and any classical post-processing. In this way we create the dataset used to train the machine learning models. Further details of the output of the quantum runs are given in \autoref{sup:sqdrift_runs}, and the details of the SqDRIFT algorithm and its parameters in \autoref{sec:sup_sqdrift}.

\subsection{Training the encoder}\label{sec:methodology_train_enc}

For physical systems, and specifically for eigenvalue problems arising in Hamiltonian simulations, a sensible choice for representing operators is their Pauli decomposition (with appropriate padding as required). While in general the number of terms in the decomposition scales as $4^{Q} = N^{2}$, i.e.\ with the number of entries in the matrix, in practice sparsity and symmetries considerably reduce the number of terms required. 

In general, the Pauli decomposition of an operator $H \in \mathbb{C}^{N \times N}$ can be expressed as                    
\begin{equation}                                                  
H = \sum_{i=1}^{T} c_i P_i,                 
\end{equation}

where $T$ is the number of Pauli terms, $c_i \in \mathbb{R}$ are scalar coefficients, and $P_i \in \{I, X, Y, Z\}^{\otimes Q}$ are tensor-product Pauli strings (e.g.\ $XIZ\cdots Z$), with $Q$ the number of qubits required to represent the Hamiltonian. 
We train a Pauli encoder $\mathcal{E}^{\mathrm{iso}}_{\gamma}\colon H \to \mathbb{R}^{d}$, parametrized by $\gamma$, that projects the operator into a $d$-dimensional space ($d = 768$), with embeddings constrained to the unit sphere by $\ell_2$ normalization. The encoder is trained such that operators sharing the same spectrum (isospectral operators) are assigned similar embeddings, while operators with different spectra are mapped far apart. More details about the encoder architecture can be found in \autoref{sec:sup_encoder_architecture}.

During training we use a contrastive objective over triplets of operators: an anchor Hamiltonian $A_n$, a positive Hamiltonian $P_n$ drawn from the same isospectral family as the anchor, and a negative Hamiltonian $N_n$ from a different family. 

The encoder produces embeddings
\[
    \mathbf{a}_n = \mathcal{E}^{\mathrm{iso}}_\gamma(A_n), \quad
    \mathbf{p}_n = \mathcal{E}^{\mathrm{iso}}_\gamma(P_n), \quad
    \mathbf{n}_n = \mathcal{E}^{\mathrm{iso}}_\gamma(N_n),
\]
 and we minimize a margin-based contrastive loss~\autocite{hadsell2006contrastive} applied to all three pairs within each triplet~\autocite{schroff2015facenet},
\begin{equation}
    \ \mathcal{L}(\gamma) = \frac{1}{N}\sum_{n=1}^{N} \Big[d(\mathbf{a}_n,\mathbf{p}_n)^2                         + \max\!\big(1 - d(\mathbf{a}_n,\mathbf{n}_n),\, 0\big)^2   + \max\!\big(1 - d(\mathbf{p}_n,\mathbf{n}_n),\, 0\big)^2   \Big],
\end{equation}
 where $d(\cdot,\cdot)$ is the cosine distance and $N$ is the number of triplets in a batch. Attractive terms pull isospectral pairs together, while the hinged repulsive terms push pairs with distinct spectra apart until their cosine distance reaches unity.

\subsection{Training the models for runtime and error prediction}\label{sec:methodology_training_the_models}

\textbf{Synthetic Hamiltonians}
For each matrix in the synthetic Hamiltonians dataset and each $m$--$\Delta t$ configuration from the grid, we train independent regressors for the runtime $\hat{t}$ and the state-propagation error $\epsilon_{prop}$. For training, we quantile-transform the error to stabilize its heavy-tailed target distribution and log-transform the runtime. The models use as features the configurations together with the 768-dimensional embeddings, reduced to 50 dimensions using PCA~\autocite{pearson1901pca,sklearn_api}. Both regressors are random forests. We fit the models on the training part of the dataset, which contains Hamiltonians of sizes $N=2^3$ to $N=2^7$, and evaluate them on Hamiltonians of sizes up to $N=2^{10}$, holding out sizes $N=2^8$, $N=2^9$, and $N=2^{10}$ from training; this yields in-distribution results for matrices up to size $N=2^7$ and out-of-distribution (OOD) results for sizes $N=2^8$, $N=2^9$, and $N=2^{10}$. To evaluate performance we use the Spearman correlation coefficient between true and predicted runtimes and errors, and the median absolute percentage error (APE) computed as:  
\begin{equation*}
    \mathrm{med.\ APE} = \operatorname{median}_{i=1,\dots,N}
  \left( \frac{|y_{\mathrm{true}_i} - y_{\mathrm{pred}_i}|}{|y_{\mathrm{true}_i}|} \times 100\% \right).
\end{equation*}

We use the best models to select the $m$ and $\Delta t$ configurations for the downstream SKQD simulations and to compare the results. \autoref{tab:methods-panel-synthetic} lists all the feature combinations we try, and the detailed results of all combinations are given in \autoref{sup:skqd_simulations}.

\begin{table}[htbp]
  \centering
  \caption{\textbf{Feature representations and estimators compared on the synthetic Hamiltonians dataset.} Every representation is concatenated with the two configuration parameters, the Krylov subspace dimension $m$ and the propagation timestep $\Delta t$. The encoder embedding is reduced by PCA fitted on the training split only.}
  \label{tab:methods-panel-synthetic}
  \small
  \begin{tabular}{@{}llrl@{}}
    \toprule
    \textbf{Feature set} & \textbf{Composition} & \textbf{Dimension} & \textbf{Estimator} \\
    \midrule
    Baseline
      & $m$, $\Delta t$
      & 2   & Random forest \\
    Matrix features
      & 130 matrix descriptors $+$ $m$, $\Delta t$
      & 132 & Random forest \\
    Encoder features
      & PCA-50 of the 768-dim embedding $+$ $m$, $\Delta t$
      & 52  & Random forest \\
    \bottomrule
  \end{tabular}
\end{table}

\textbf{Molecular Hamiltonians}\label{sec:methods_molecular_hamiltonians}
For each of the 50 selected matrices and the 12 SqDRIFT configurations, we predict the runtime $\hat{t}$ and the error $\hat{\epsilon}$. We use the full 768-dimensional embeddings of the Hamiltonians from the encoder, together with the configurations, as features to train a random forest regressor for the runtime $\hat{t}$ and the 8 PCA-reduced features from the embeddings to train an Ordinary Least Squares model for the ground state energy approximation error $\hat{\epsilon}$. We log-transform the targets. We split the dataset into three sets: train, in-distribution (ID) test, and out-of-distribution (OOD) test. The train set contains 34 molecules of 20 to 40 qubits, the ID test set contains 11 molecules of 20 to 40 qubits, and the OOD test set contains 5 larger molecules of 42 to 50 qubits. This split tests the generalizability of the models to matrices of sizes beyond those seen during training. \autoref{tab:methods-panel-molecular} lists the model and feature combinations used to learn the targets, and further details of the results are given in \autoref{sup:ablations_molecular_hamiltonians}.

\begin{table}[htbp]
  \centering
  \caption{Feature representations and estimators compared on the molecular Hamiltonians. The physics block holds four molecule features (\texttt{n\_electrons}, \texttt{n\_orbitals}, \texttt{n\_qubits} and
  $\log_{10}$ of the configuration-interaction dimension); the config block holds the six SqDRIFT circuit features: \texttt{num\_groups}, \texttt{num\_rand}, and four entries (\texttt{t1}--\texttt{t4}) representing a fixed-length encoding of the evolution-time list. The two time configurations considered are \texttt{[1,2,3]} and \texttt{[1,2,3,4]}; the former is padded to \texttt{[1,2,3,0]} so that both configurations have the same dimensionality. In both cases, t1-t3 are the same, thus reducing the number of effective variables to 3 informative ones (\texttt{num\_groups}, \texttt{num\_rand}, m), with $m \in \{3,4\}$ being the Krylov dimension and $\Delta t=1$.
  The 768-dimensional Hamiltonian embedding enters either in full or reduced by
  PCA to eight components, fitted on the training split only. Every regression
  feature set includes the config block; the classification panel additionally
  fits two feature sets that isolate it.}
  \label{tab:methods-panel-molecular}
  \small
  \begin{tabular}{@{}llr@{}}
    \toprule
    Feature set & Composition & Dim. \\
    \midrule
    \multicolumn{3}{@{}l}{\emph{Regression} --- OLS, Ridge, random forest,
      XGBoost~\autocite{chen2016xgboost}, LightGBM~\autocite{ke2017lightgbm}} \\[2pt]
    Physics              & 4 descriptors $+$ config              & 10  \\
    Emb-PCA8             & PCA-8 of the embedding $+$ config     & 14  \\
    Emb-full             & 768 embedding dims $+$ config         & 774 \\
    Physics $+$ Emb-PCA8 & 4 descriptors $+$ PCA-8 $+$ config    & 18  \\
    \midrule
    \multicolumn{3}{@{}l}{\emph{Classification} --- logistic regression, random
      forest, XGBoost~\autocite{chen2016xgboost}, LightGBM~\autocite{ke2017lightgbm}, RBF-SVM, OLS-CDF} \\[2pt]
    Config                        & config only                        & 6   \\
    Physics                       & 4 descriptors only                 & 4   \\
    Physics $+$ config            & 4 descriptors $+$ config           & 10  \\
    Emb-PCA8 $+$ config           & PCA-8 of the embedding $+$ config  & 14  \\
    Physics $+$ Emb-PCA8 $+$ config & 4 descriptors $+$ PCA-8 $+$ config & 18  \\
    Emb-full $+$ config           & 768 embedding dims $+$ config      & 774 \\
    \addlinespace[2pt]
    Majority, Prevalence          & no features                        & 0   \\
    \bottomrule
  \end{tabular}
\end{table}

\section*{Acknowledgments}
D.G., F.Z. acknowledge the funding by NCCR Catalysis (grant number 225147), a National Centre of Competence in Research funded by the Swiss National Science Foundation. This work was supported by a grant from the Swiss National Supercomputing Centre (CSCS) under project ID PJ05 on the IBM Quantum Cloud. The authors would also like to thank Dr. Julian Schuhmacher and Samuele Piccinelli for the insightful discussions on SKQD and SqDRIFT circuits.

\section*{Data availability}
Data for the molecular Hamiltonians (molecule and integrals) as well as samples collected from the QPU runs are available to aid experiment reproducibility~\autocite{molecules_for_sqd_benchmarking}.

\clearpage






\printbibliography
\clearpage

\setcounter{section}{0}
\renewcommand{\thesection}{S\arabic{section}}
\renewcommand{\thesection}{\thesection.\arabic{section}}

\setcounter{figure}{0}
\renewcommand{\thefigure}{S\arabic{figure}}

\setcounter{table}{0}
\renewcommand{\thetable}{S\arabic{table}}

\setcounter{equation}{0}
\renewcommand{\theequation}{S\arabic{equation}}

\begin{appendices}
\section{The SKQD Algorithm}\label{sec:sup_skqd}
Sample-based Krylov Quantum Diagonalization (SKQD)~\autocite{yu_quantum-centric_2025} is a quantum diagonalization algorithm that uses bitstrings sampled from quantum states to build a subspace, into which the original Hamiltonian $H$ is subsequently projected and diagonalized. SKQD is based on quantum-selected configuration interaction (QSCI)~\autocite{kanno2023qsci} and its quantum-centric-supercomputing realization with self-consistent configuration
recovery~\autocite{robledomoreno2025sqd}, and belongs to the broader family of quantum subspace methods~\autocite{motta2024subspace}. The quantum state evolves from a reference state over multiple intervals. The two key parameters, on which this work also focuses, are $m$, the Krylov subspace dimension, and $\Delta t$, the timestep for the evolution of the reference state. The evolution of the reference state $\psi_0$ that determines the Krylov states is given by \autoref{eq:sup_sqkd_time_evol}:
\begin{equation}
    |\psi_k\rangle = e^{-i k\Delta t H} |\psi_0\rangle
    \label{eq:sup_sqkd_time_evol}
\end{equation}
where $k \in \{0,\dots,m-1\}$, $m$ is the Krylov subspace dimension, and $H$ is the Hamiltonian whose ground state is to be computed.
The outline of the SKQD algorithm~\autocite{yu_quantum-centric_2025}, is as follows:
\begin{enumerate}
    \item Start from a reference state $|\psi_0\rangle$.
    \item For each $k \in \{0, ... , m-1\}$, create $M$ copies of $\psi_k$.
    \item Measure each of the copies of $\psi_k$ in the computational basis to obtain a set of bitstrings $\mathcal{X}_k$.
    \item Project $H$ into the subspace spanned by all unique bitstrings from all sets $\mathcal{X}_k$ to obtain $\bm{\hat{H}}$.
    \item Diagonalize $\bm{\hat{H}}$ classically to find the approximate ground state $\bm{\Psi}$.
\end{enumerate}

SKQD converges under two assumptions: (i) a polynomial overlap of the reference state $\psi_0$ with the true ground state $\bm{\Psi}$, and (ii) sparsity of the true ground state \autocite{yu_quantum-centric_2025}. Given that these assumptions hold, the approximation quality depends on the choice of the parameters $m$ and $\Delta t$. In theory, a large enough $m$ and a small enough $\Delta t$ recover the true ground state; in practice, however, this is infeasible, as the resulting large circuit depths produce noisy results on Noisy Intermediate-Scale Quantum (NISQ) devices. We therefore focus on models that reliably predict the \texttt{error} and \texttt{runtime} for a Hamiltonian $H$ given a grid of $m$ and $\Delta t$ values.

\section{The SqDRIFT Algorithm} \label{sec:sup_sqdrift}
While SKQD avoids variational optimization and produces reliable ground state approximations when the initial assumptions hold, it generates deep circuits that yield poor approximations on NISQ devices for dense Hamiltonians. SqDRIFT\autocite{piccinelli_quantum_2026} addresses this issue by combining the Krylov subspace idea from SKQD\autocite{yu_quantum-centric_2025} with the randomization idea from the qDRIFT algorithm\autocite{campbell_random_2019}. To circumvent the problems arising from running deep circuits on NISQ devices, SqDRIFT instead runs multiple realizations of smaller circuits. Following \autocite{piccinelli_quantum_2026}, SqDRIFT works as follows:
\begin{enumerate}
    \item Given the Pauli decomposition of a Hamiltonian $H = \sum_{j=1}^{M} c_j P_j$, where $c_j$ is the coefficient of the Pauli string $P_j$, and $P_j= \bigotimes_{i=1}^{n} \sigma_i$ is the tensor product of the Pauli operators $\sigma_i \in \{I,X,Y,Z\}$.
    \item From all $M$ Pauli terms and coefficients sample \texttt{num\_groups} randomly with probability $p_j= |c_j| / \lambda$, where $\lambda = \sum_{j=1}^{M} |c_j|$. The Hamiltonian is then represented by a subset of Pauli terms:
    \begin{equation}
        \tilde{H}=\lambda \sum_{j=1}^{\texttt{num\_groups}} p_j \operatorname{sgn}(c_j)P_j 
    \end{equation}
    The time evolution under $\tilde{H}$ is then represented as a product of single term exponentials:
    \begin{equation}
        U_{\mathrm{qDRIFT}} = \prod_{j=1}^{\texttt{num\_groups}} e^{-i\,\operatorname{sgn}(h_{j})\,P_{j}\,\lambda t / \texttt{num\_groups}}
    \end{equation}
    \item Repeat step 2 \texttt{num\_rand} times. This yields \texttt{num\_rand} representations $\tilde{H}$ of $H$, which are transpiled to quantum circuits.
    \item The timestep $\Delta t$ controls the evolution of the reference state $\psi_0$ for each of these realizations from step 3:
    \begin{equation}
        |\psi_k\rangle = e^{-i k\Delta t \tilde{H}} |\psi_0\rangle
    \label{eq:sqkd_time_evol}
    \end{equation}
    $\Delta t = 1$ is used throughout the experiments, with the Krylov dimension $m\in\{3,4\}$.
    \item Measure the $\psi_k$ in the computational basis and construct a subspace $\mathcal{B}$ spanned by all unique bitstrings returned by the measurements.
    \item Project the original $H$ onto $\mathcal{B}$ and diagonalize it to find the ground state $\Psi$.
\end{enumerate}

By creating multiple shallower circuits, SqDRIFT circumvents the issues introduced by SKQD. As in SKQD, the choice of parameters is important for the quality of the ground state approximation. In this work, we focus on predicting the \texttt{runtime} and ground state energy \texttt{error} for a specific Hamiltonian given \texttt{num\_groups}, \texttt{num\_rand}, and $\Delta t$.

\section{Synthetic Hamiltonian Generation}\label{sec:sup_synthetic_ham}

The encoder architecture must be paired with a dataset that enables training at scale. Since data availability is one of the key requirements for training AI models, we construct a dataset composed of a series of isospectral operators that exhibit diverse Pauli-sparsity patterns and structural variability while avoiding uncontrolled exponential growth in the number of Pauli terms. The goal is not structured diagonalization, but rather to sample families of matrices that share the same eigenvalue spectrum while appearing ``random'' for downstream machine learning applications.

We employ two complementary mechanisms: (1) isospectral trajectories generated via adjoint-unitary evolution in the Pauli basis; (2) synthetic scaling through block-diagonal composition followed by global mixing.

\subsection{Data generation for isospectral operator trajectories}
\label{sec:sup_data_gen_isospectral_traj}

For any given operator, one \emph{isospectral trajectory} can be obtained by iteratively conjugating a sparse Pauli expansion with single-Pauli generators. This produces operators with varying Pauli support while preserving their eigenvalue spectrum.

\begin{enumerate}
    \item \textbf{Initialization and sparse truncation}
    Let $H \in \mathbb{C}^{N \times N}$ be a random Hermitian matrix. Its Pauli decomposition is
\[
H = \sum_{i=1}^{T} c_i P_i,
\]
where \(P_i \in \{I, X, Y, Z\}^{\otimes Q}\). To define a controlled initial sparsity, we retain 
only a subset $\mathcal{S}$ of $k$ terms, selected by a criterion such as:
\begin{itemize}
    \item largest-magnitude coefficients $|c_j|$,
    \item uniform random sampling,
    \item weighted sampling by coefficient magnitude.
\end{itemize}

The truncated operator is
\[
H_0 = \sum_{i \in \mathcal{S}} c_i P_i,
\]
and serves as the initial point of the trajectory.

\item \textbf{Isospectral Evolution via Pauli Conjugation}
We generate the trajectory by successive adjoint actions
\[
H_{m+1} = e^{i\phi_m G_m}\, H_m\, e^{-i\phi_m G_m},
\]
where $G_m$ is a Pauli string chosen to anticommute with the currently largest-magnitude 
off-diagonal term in $H_m$. These transformations preserve the hermiticity and the spectrum of the operator while modifying the Pauli-sparsity pattern and remaining within the Pauli manifold.
For any Pauli string $P$ appearing in $H_m$, its update under the rotation generated by $G$ is:
\[
e^{i\phi G} P e^{-i\phi G} =
\begin{cases}
P, & [G,P]=0, \\[4pt]
\cos(2\phi)\, P \;-\; \sin(2\phi)\,(iGP), & \{G,P\}=0.
\end{cases}
\]
When $\{G,P\}=0$, the operator $iGP$ is again a Pauli string, so the evolution remains 
within the Pauli manifold.

\item \textbf{Complexity Control}
To avoid uncontrolled proliferation of Pauli terms, we use: (1) angle damping for $\phi_m$, (2) pruning of small-magnitude coefficients, (3) an early-stopping condition based on (a) the number of Pauli terms exceeding a threshold, or (b) the magnitude of off-diagonal contributions falling below~$\epsilon$.

The resulting trajectory:
\[
\bigl\{ H_0,\, H_1,\, H_2,\, \dots \bigr\}
\]
consists of isospectral operators with varied sparsity and structure.
\end{enumerate}

\subsection{Synthetic Scaling via Block Composition and Mixing}\label{sec:sup_data_scaling}

To construct high-dimensional operators (e.g., $N >> 2^{11}$ or $Q>11$), we can combine smaller 
operators into block structures and then introduce interactions through global rotations.

\begin{enumerate}
    \item \textbf{Block-Diagonal Composition} 
    Given two operators $H_A$ and $H_B$ with sizes $N_A = 2^{Q_A}$ and $N_B = 2^{Q_B}$ respectively, we can construct a block-diagonal operator of size $N_{\mathrm{comp}} = 2^{Q_A + Q_B + 1}$
\[
H_{\mathrm{comp}} = H_A \oplus H_B
\]
This can be written in the Pauli basis as
\[
H_{\mathrm{comp}}
= \frac{1}{2} \left[
I \otimes (H_A + H_B)
+ Z \otimes (H_A - H_B)
\right].
\]
Note that this construction guarantees that we can determine the spectrum of the newly created operator by leveraging those of the two combined blocks. This enables scaling to high dimensions while retaining knowledge of the operator spectrum.
\item \textbf{Global Mixing Rotation}
To introduce coupling while preserving spectra, we apply a unitary rotation on the newly obtained operator:
\[
H_{\mathrm{mixed}} 
= e^{i\phi X\_0}\, H_{\mathrm{comp}}\, e^{-i\phi X\_0},
\]
where $X\_0 = X \otimes I^{\otimes Q}$. This breaks the explicit block structure and produces 
a higher-dimensional operator with a more entangled Pauli structure, yet still isospectral to $H_{\mathrm{comp}}$.
\item \textbf{Global Mixing via a Chain of Random Unitary Rotations}
In addition, we can generalize the single rotation approach, as we can apply a sequence of unitaries
\[
U_k = e^{i \phi_k G_k},
\]
where $\phi_k$ are random mixing angles and $G_k$ are randomly selected
Pauli strings of varying locality (from 1-local to $Q$-local).

The total mixing unitary is
\[
U_{\mathrm{mix}} = U_K U_{K-1} \cdots U_1,
\]
and the mixed operator is obtained through conjugation:
\[
H_{\mathrm{mixed}} 
= U_{\mathrm{mix}}\, H_{\mathrm{comp}}\, U_{\mathrm{mix}}^\dagger.
\]

The single-rotation case used previously,
$e^{i\phi X_0}$,
is recovered by choosing $G_k = X_0$ for all $k$.
The general construction, however, allows $G_k$ to be any Pauli
string, enabling richer mixing while preserving spectral invariance.
\end{enumerate}

\section{Molecular Hamiltonians and SqDRIFT circuit definition}\label{sec:sup_chemistry_ham}

We generate the 50 molecular Hamiltonians from molecular geometries provided as \texttt{.xyz} files, using a four-stage pipeline implemented with the PySCF~\autocite{sun2020pyscf} package.

\begin{enumerate}
    \item \textbf{Molecular setup.} We derive the net charge from the SMILES string of each molecule via RDKit~\autocite{rdkit} and pass the molecular structure, including the charge information, to PySCF~\autocite{sun2020pyscf} with the cc-pVDZ~\autocite{dunning1989ccpvdz} basis set.

    \item \textbf{Mean-field reference.} We perform a Restricted Hartree--Fock (RHF) calculation to convergence with a tolerance of $10^{-10}$ for a maximum of 300 cycles to obtain the reference molecular-orbital coefficients.

    \item \textbf{MP2 natural orbitals.} We run second-order M{\o}ller--Plesset perturbation theory (MP2)~\autocite{jensen1988mp2no} on the RHF reference. We transform the one-particle reduced density matrix (1-RDM) to the atomic-orbital basis and diagonalize it to yield natural orbitals sorted by descending Natural Orbital Occupation Numbers (NOONs).

    \item \textbf{Active-space selection and FCIDUMP generation.} We select active spaces by retaining all natural orbitals whose NOON differes from the fully occupied (2.00) or fully unoccupied (0.00) values by more than 0.02, which identifies orbitals with fractional occupation indicative of strong correlation. We always include, in addition, the top-five occupied and lowest-five virtual states to guarantee the presence of orbitals around the HOMO--LUMO gap. We transform the one- and two-electron integrals over the selected MP2~\autocite{jensen1988mp2no} natural orbitals into the active-space basis and write them to disk in the FCIDUMP format.
\end{enumerate}

We then convert each second-quantized active-space Hamiltonian stored in the FCIDUMP into a qubit Hamiltonian via the Jordan--Wigner transformation~\autocite{jordan1928wigner}, mapping each spin orbital to one qubit. The resulting Pauli decomposition serves as direct input to the SqDRIFT circuits, following the fermionic sampling protocol~\autocite{piccinelli_quantum_2026}.

\autoref{tab:molecules} lists the 50 molecules with their dataset split, chemical class, molecular formula, active-space size, and SqDRIFT two-qubit gate depth.

\begingroup
\small
\setlength{\tabcolsep}{3pt}
\begin{longtable}{@{}llp{0.21\linewidth}p{0.1\linewidth}rrrrr@{}}
\caption{\textbf{The 50 molecular Hamiltonians.} Columns: dataset split; global molecule index; chemical class; molecular formula; number of active-space spatial orbitals ($N_\text{orb}$); number of active electrons ($N_\text{elec}$); number of qubits ($N_q$); mean SqDRIFT two-qubit gate depth with standard deviation; Truncated Davidson ground-state energy estimate (Hartree). The estimated error of this reference energy is between $10^{-6}$ and $10^{-4}$, thus below chemical accuracy at $1.6\times 10^{-3}$ Ha.}
\label{tab:molecules}\\
\toprule
\textbf{Split} & \textbf{Idx} & \textbf{Class} & \textbf{Formula} & $N_\text{orb}$ & $N_\text{elec}$ & $N_q$ & \textbf{Circuit depth} & \textbf{TD GS est. (Ha)} \\
\midrule
\endfirsthead
\multicolumn{9}{c}{\tablename~\thetable{} (continued)}\\
\toprule
\textbf{Split} & \textbf{Idx} & \textbf{Class} & \textbf{Formula} & $N_\text{orb}$ & $N_\text{elec}$ & $N_q$ & \textbf{Circuit depth} & \textbf{TD GS est. (Ha)} \\
\midrule
\endhead
\midrule
\multicolumn{9}{r}{\emph{Continued on next page}}\\
\endfoot
\bottomrule
\endlastfoot
Train & 30  & Alkanes and cycloalkanes         & C$_4$H$_8$       & 16 & 22 & 32 & $716 \pm 237$  & $-156.180$ \\
Train & 51  & Alkenes and dienes                & C$_2$H$_4$       & 10 & 10 & 20 & $492 \pm 188$  & $-78.129$ \\
Train & 63  & Alkenes and dienes                & C$_5$H$_8$       & 16 & 22 & 32 & $489 \pm 193$  & $-194.074$ \\
Train & 101 & Alkynes                           & C$_3$H$_4$       & 12 & 14 & 24 & $317 \pm 139$  & $-115.994$ \\
Train & 102 & Alkynes                           & C$_4$H$_6$       & 14 & 18 & 28 & $354 \pm 148$  & $-155.021$ \\
Train & 106 & Alkynes                           & C$_4$H$_6$       & 14 & 18 & 28 & $547 \pm 202$  & $-155.033$ \\
Train & 136 & Alkynes                           & C$_3$H$_2$O$_2$  & 17 & 20 & 34 & $484 \pm 200$  & $-264.676$ \\
Train & 142 & Alkynes                           & C$_2$H$_2$       & 10 & 10 & 20 & $429 \pm 132$  & $-76.965$ \\
Train & 151 & Arenes and heteroarenes           & C$_6$H$_6$       & 19 & 24 & 38 & $858 \pm 235$  & $-230.873$ \\
Train & 175 & Arenes and heteroarenes           & C$_4$H$_4$N$_2$  & 20 & 24 & 40 & $786 \pm 309$  & $-262.923$ \\
Train & 204 & Alcohols and phenols              & C$_4$H$_{10}$O   & 19 & 26 & 38 & $626 \pm 251$  & $-232.277$ \\
Train & 253 & Ethers and acetals                & C$_4$H$_{10}$O   & 18 & 26 & 36 & $465 \pm 209$  & $-232.258$ \\
Train & 301 & Aldehydes and ketones             & C$_2$H$_4$O      & 12 & 14 & 24 & $335 \pm 131$  & $-153.057$ \\
Train & 371 & Carboxylic acids                  & C$_3$H$_4$O$_2$  & 17 & 22 & 34 & $485 \pm 187$  & $-265.852$ \\
Train & 377 & Carboxylic acids                  & C$_2$HNO$_2$     & 17 & 20 & 34 & $644 \pm 324$  & $-280.754$ \\
Train & 379 & Carboxylic acids                  & C$_2$H$_4$O$_3$  & 19 & 24 & 38 & $615 \pm 229$  & $-302.888$ \\
Train & 381 & Carboxylic acids                  & C$_2$H$_3$BrO$_2$& 18 & 24 & 36 & $513 \pm 225$  & $-2799.796$ \\
Train & 501 & Amines                            & CH$_5$N          & 11 & 12 & 22 & $352 \pm 145$  & $-95.312$ \\
Train & 507 & Amines                            & C$_3$H$_9$N      & 16 & 22 & 32 & $507 \pm 174$  & $-173.383$ \\
Train & 526 & Amines                            & C$_4$H$_{11}$N   & 18 & 26 & 36 & $469 \pm 192$  & $-212.410$ \\
Train & 562 & Nitriles                          & C$_5$H$_9$N      & 20 & 28 & 40 & $546 \pm 257$  & $-249.198$ \\
Train & 580 & Nitriles                          & C$_3$H$_6$N$_2$  & 17 & 22 & 34 & $487 \pm 224$  & $-226.156$ \\
Train & 603 & Nitro compounds and N-oxides      & C$_3$H$_7$NO$_2$ & 20 & 28 & 40 & $599 \pm 309$  & $-321.956$ \\
Train & 652 & Sulfur compounds                  & C$_2$H$_6$S      & 13 & 16 & 26 & $476 \pm 220$  & $-476.833$ \\
Train & 659 & Sulfur compounds                  & C$_3$H$_8$S      & 15 & 20 & 30 & $424 \pm 174$  & $-515.867$ \\
Train & 661 & Sulfur compounds                  & C$_4$H$_{10}$S   & 17 & 24 & 34 & $479 \pm 196$  & $-554.913$ \\
Train & 662 & Sulfur compounds                  & C$_4$H$_{10}$S   & 18 & 26 & 36 & $500 \pm 230$  & $-554.911$ \\
Train & 669 & Sulfur compounds                  & C$_2$H$_6$S      & 13 & 16 & 26 & $562 \pm 187$  & $-476.826$ \\
Train & 684 & Sulfur compounds                  & C$_5$H$_{12}$S   & 20 & 30 & 40 & $556 \pm 259$  & $-593.944$ \\
Train & 701 & Charged and zwitterionic          & CH$_6$N$^+$      & 11 & 12 & 22 & $367 \pm 163$  & $-95.675$ \\
Train & 707 & Charged and zwitterionic          & C$_5$H$_{14}$N$^+$& 20 & 30 & 40 & $475 \pm 233$  & $-251.845$ \\
Train & 732 & Charged and zwitterionic          & C$_2$H$_8$NO$^+$ & 15 & 20 & 30 & $569 \pm 256$  & $-209.596$ \\
Train & 734 & Charged and zwitterionic          & CH$_3$O$_3$S$^-$ & 18 & 24 & 36 & $792 \pm 284$  & $-661.873$ \\
Train & 741 & Charged and zwitterionic          & C$_4$H$_7$O$_2^-$& 20 & 28 & 40 & $665 \pm 321$  & $-305.468$ \\
\midrule
Test ID & 1   & Alkanes and cycloalkanes        & CH$_4$           & 10 & 10 & 20 & $296 \pm 120$  & $-40.286$ \\
Test ID & 2   & Alkanes and cycloalkanes        & C$_2$H$_6$       & 11 & 12 & 22 & $436 \pm 148$  & $-79.306$ \\
Test ID & 4   & Alkanes and cycloalkanes        & C$_4$H$_{10}$    & 16 & 22 & 32 & $558 \pm 192$  & $-157.368$ \\
Test ID & 52  & Alkenes and dienes              & C$_3$H$_6$       & 12 & 14 & 24 & $376 \pm 147$  & $-117.167$ \\
Test ID & 60  & Alkenes and dienes              & C$_4$H$_6$       & 14 & 18 & 28 & $654 \pm 243$  & $-155.039$ \\
Test ID & 68  & Alkenes and dienes              & C$_6$H$_8$       & 19 & 26 & 38 & $854 \pm 268$  & $-231.968$ \\
Test ID & 575 & Nitriles                        & C$_2$H$_3$NO     & 15 & 18 & 30 & $498 \pm 189$  & $-206.977$ \\
Test ID & 643 & Nitro compounds and N-oxides    & C$_2$H$_4$BrNO$_2$& 20 & 28 & 40 & $534 \pm 257$  & $-2854.726$ \\
Test ID & 702 & Charged and zwitterionic        & C$_2$H$_8$N$^+$  & 13 & 16 & 26 & $399 \pm 178$  & $-134.710$ \\
Test ID & 717 & Charged and zwitterionic        & C$_2$H$_5$NO$_2$ & 18 & 24 & 36 & $591 \pm 245$  & $-282.983$ \\
Test ID & 724 & Charged and zwitterionic        & C$_3$H$_5$O$_2^-$& 17 & 24 & 34 & $630 \pm 315$  & $-266.420$ \\
\midrule
Test OOD & 7   & Alkanes and cycloalkanes       & C$_7$H$_{16}$    & 25 & 38 & 50 & $597 \pm 224$  & $-274.497$ \\
Test OOD & 315 & Aldehydes and ketones          & C$_6$H$_{12}$O   & 24 & 34 & 48 & $494 \pm 213$  & $-309.230$ \\
Test OOD & 591 & Nitriles                       & C$_5$H$_8$ClN    & 23 & 32 & 46 & $474 \pm 233$  & $-708.138$ \\
Test OOD & 665 & Sulfur compounds               & C$_5$H$_{12}$S   & 21 & 30 & 42 & $521 \pm 208$  & $-593.963$ \\
Test OOD & 713 & Charged and zwitterionic       & C$_3$H$_6$NO$_2^-$& 22 & 28 & 44 & $577 \pm 222$  & $-321.525$ \\
\end{longtable}
\endgroup

\section{Encoder Architecture}\label{sec:sup_encoder_architecture}

The model architecture is an encoder-only transformer that processes each Pauli
decomposition of $H$ as a sequence of tokens whose length grows linearly in the
number of terms $T$ and qubits $Q$. It consumes three types of input features:
global features (system-level summary statistics of the coefficient distribution) $g \in \mathbb{R}^{10}$; per-qubit features (per-dimension state information, including a normalized axis histogram over the I/X/Y/Z operators, mean log-coefficient magnitude, normalized position, and term count) $q_i \in \mathbb{R}^{7}$; and per-term features (Pauli string representations with coefficient magnitude and phase information) $t_i \in \mathbb{R}^{9}$. The encoder embeds these heterogeneous inputs into a unified hidden space through learned linear projections, each followed by layer normalization and a learned additive type embedding, and concatenates them into the sequence [g, q$_1$, ..., q$_Q$, t$_1$, ..., t$_T$].

The encoder is a two-stage Multiscale Byte Language Model (MBLM)\autocite{egli_multiscale_2025} at hidden size 768: a global stage of 5 transformer-encoder layers over up to 2048 coarse positions, followed by a local stage of 1 layer over 100 inner positions. Sequence representations are obtained by mask-aware mean pooling applied separately to the qubit and term segments, so that padding positions introduced by dynamic batching do not contribute to the pooled vectors. These are concatenated with the global token's hidden state, rescaled by learned per-type scalars, and passed through a projection MLP to a 768-dimensional embedding, which is $L_2$-normalized. Training uses a purely contrastive (triplet) objective on cosine distances between anchor, positive, and negative operators; the auxiliary reconstruction head is disabled ($\alpha = 0$). We optimize in BF16 with a learning rate of $1 \times 10^{-4}$ and weight decay 0.01.


\section{Creating the synthetic Hamiltonians dataset, ablations of the models and the classical SKQD simulations}\label{sup:skqd_simulations}

For each synthetic Hamiltonian, we form 48 configurations from the Cartesian product of $m=\{5,7,10,15,20,25\}$ and $\Delta t=\{0.01, 0.02,0.05,0.1,0.2,0.5,1,2\}$. The Krylov--Arnoldi iterations produce the state-propagation error $\epsilon_{prop}$, a classical matrix-exponential error that is not the SKQD ground-state energy error, and the classical wall-clock runtime. These two quantities are the regression targets, serving as classical surrogates for the corresponding SKQD/QPU quantities. Once the models are trained, for each Hamiltonian $H$ and $m$--$\Delta t$ combination they produce the predicted state-propagation error $\hat{\epsilon}_{prop}$ and runtime $\hat{t}$.

As shown in \autoref{tab:surrogate-model-selection}, we compare three feature representations, in every case using a random forest (500 trees, maximum depth 20) trained separately for each target: the two grid parameters $(m, \Delta t)$ alone, which serve as a parameter-only baseline; 132 handcrafted matrix descriptors (further details in \autoref{tab:matrix-features}); and a 768-dimensional learned encoder embedding reduced to its first 50 principal components, the latter two both concatenated with $(m, \Delta t)$. We train each representation on two splits, one covering all matrix dimensions and one restricted to $N \le 128$, so as to probe extrapolation to matrices larger than any seen during training. Details of the number of samples per dimension and per split are given in \autoref{tab:synthetic-splits}. We use Spearman's rank correlation $\rho$ and the median absolute percentage error (APE) as evaluation metrics, since in this setting the correct ranking of the configurations matters more than an extremely precise prediction. Both learned representations are far ahead of the parameter-only baseline: on the error target they reach $\rho = 0.97$–$0.99$ against $0.66$–$0.68$ for $(m, \Delta t)$ alone, while the median APE falls from roughly $68$–$85\%$ to $12$–$20\%$. The ordering is the same for the runtime target. Trained on all dimensions, the matrix and encoder features perform almost identically, within $2.2$ percentage points of median APE and $0.004$ in $\rho$ of one another. Restricting the training matrices to $N \le 128$, while still yielding strong ranking correlations, costs both representations accuracy, and that cost falls almost entirely on the dimensions never seen during training (\autoref{fig:res_synthetic_ham_results}). This happens because the absolute magnitudes degrade considerably faster than the rankings, as a random forest cannot predict beyond the range of targets it was trained on, despite excelling at ranking the configurations. Comparing the handcrafted and encoded features, the drawback of the handcrafted descriptors is their cost: computing them scales steeply with matrix size, whereas an encoder embedding is obtained from a single forward pass. Since the two representations perform comparably in distribution, we select the encoder features, as they extend to Hamiltonians of arbitrary size.

\begin{table}[htbp]
  \centering
  \caption{\textbf{Model and feature comparison.} For each target we report the median absolute percentage error (APE) on the original scale and on the transformed scale in which the regressors were fitted (quantile-transformed error, $\log_{10}$ runtime), together with Spearman's rank correlation $\rho$ and its two-sided $p$-value. All numbers are computed on the held-out test set, pooled over all matrix dimensions $N \in \{8,\dots,1024\}$. \emph{Train} indicates the training split: \emph{all-}$N$ uses all dimensions, OOD restricts training to $N \le 128$, leaving $N \in \{256, 512, 1024\}$ entirely unseen (\autoref{tab:synthetic-splits}). All estimators are random forests (500 trees, maximum depth 20). Note that the two training variants fit their own quantile transform, so transformed-scale APE is comparable within a variant but not across them; original-scale APE and $\rho$ are comparable throughout.}
  \label{tab:surrogate-model-selection}
  \small
  \setlength{\tabcolsep}{4pt}
  \begin{tabular}{llrrrl}
    \toprule
    & & \multicolumn{2}{c}{Median APE (\%)} & & \\
    \cmidrule(lr){3-4}
    Feature set & Train & Original & Transf. & $\rho$ & $p$ \\
    \midrule
    \multicolumn{6}{l}{\emph{Target: state-propagation error}} \\[2pt]
    Baseline ($m$, $\Delta t$) & all-$N$ & 84.9 & 116.0 & 0.660 & $<0.01$ \\
    Baseline ($m$, $\Delta t$) & OOD & 68.4 & 84.7 & 0.681 & $<0.01$ \\
    Matrix features (132) & all-$N$ & 12.3 & 8.5 & 0.990 & $<0.01$ \\
    Matrix features (132) & OOD & 15.8 & 10.4 & 0.983 & $<0.01$ \\
    Encoder features (768\,$\to$\,PCA 50) & all-$N$ & 14.5 & 9.3 & 0.986 & $<0.01$ \\
    Encoder features (768\,$\to$\,PCA 50) & OOD & 19.5 & 16.8 & 0.969 & $<0.01$ \\
    \midrule
    \multicolumn{6}{l}{\emph{Target: runtime}} \\[2pt]
    Baseline ($m$, $\Delta t$) & all-$N$ & 98.1 & 21.1 & 0.802 & $<0.01$ \\
    Baseline ($m$, $\Delta t$) & OOD & 63.4 & 13.1 & 0.802 & $<0.01$ \\
    Matrix features (132) & all-$N$ & 39.4 & 8.9 & 0.939 & $<0.01$ \\
    Matrix features (132) & OOD & 50.9 & 10.7 & 0.875 & $<0.01$ \\
    Encoder features (768\,$\to$\,PCA 50) & all-$N$ & 39.9 & 9.0 & 0.938 & $<0.01$ \\
    Encoder features (768\,$\to$\,PCA 50) & OOD & 50.6 & 10.5 & 0.905 & $<0.01$ \\
    \bottomrule
  \end{tabular}
\end{table}

\begin{table}[htbp]
  \centering
  \caption{\textbf{Composition of the synthetic-Hamiltonian dataset by matrix dimension $N$}. \emph{all-}$N$ covers every dimension, while OOD restricts the training matrices to $N \le 128$, leaving the three largest dimensions unseen during training. Both variants are evaluated on the same test split. Training and test matrices are disjoint.}
  \label{tab:synthetic-splits}
  \small
  \setlength{\tabcolsep}{4pt}
  \begin{tabular}{rrrrrrrrr}
    \toprule
    & \multicolumn{4}{c}{all-$N$} & \multicolumn{4}{c}{OOD} \\
    \cmidrule(lr){2-5}\cmidrule(lr){6-9}
    & \multicolumn{2}{c}{Train} & \multicolumn{2}{c}{Test}
    & \multicolumn{2}{c}{Train} & \multicolumn{2}{c}{Test} \\
    \cmidrule(lr){2-3}\cmidrule(lr){4-5}\cmidrule(lr){6-7}\cmidrule(lr){8-9}
    $N$ & Samples & Matrices & Samples & Matrices & Samples & Matrices & Samples & Matrices \\
    \midrule
    8 & 54\,504 & 3\,535 & 15\,740 & 1\,010 & 54\,504 & 3\,535 & 15\,740 & 1\,010 \\
    16 & 101\,472 & 3\,469 & 28\,117 & 959 & 101\,472 & 3\,469 & 28\,117 & 959 \\
    32 & 118\,592 & 2\,750 & 34\,235 & 791 & 118\,592 & 2\,750 & 34\,235 & 791 \\
    64 & 104\,851 & 2\,496 & 27\,914 & 664 & 104\,851 & 2\,496 & 27\,914 & 664 \\
    128 & 61\,230 & 1\,467 & 18\,976 & 452 & 61\,230 & 1\,467 & 18\,976 & 452 \\
    256 & 58\,978 & 1\,420 & 15\,302 & 368 & -- & -- & 15\,302 & 368 \\
    512 & 56\,243 & 1\,362 & 14\,426 & 349 & -- & -- & 14\,426 & 349 \\
    1024 & 51\,658 & 1\,252 & 17\,021 & 412 & -- & -- & 17\,021 & 412 \\
    \midrule
    \textbf{Total} & \textbf{607\,528} & 17\,751 & \textbf{171\,731} & 5\,005
    & \textbf{440\,649} & 13\,717 & \textbf{171\,731} & 5\,005 \\
    \bottomrule
  \end{tabular}
\end{table}

\begin{figure}[htbp]
    \includegraphics[width=\linewidth]{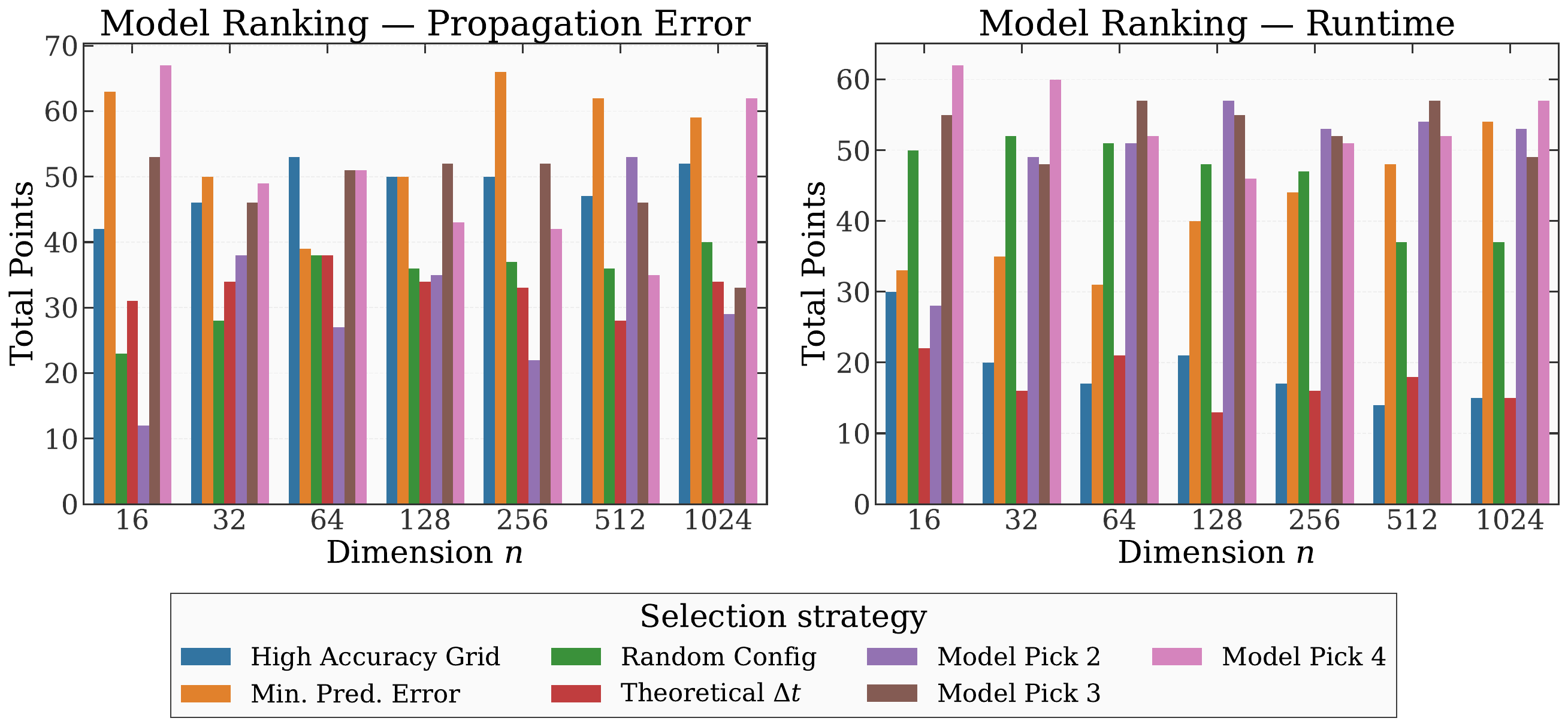}
    \caption{\textbf{Borda counts from the classical SKQD simulations.} The first plot ranks the configurations, per dimension, by the state-propagation error $\epsilon_{prop}$ of the runs. The second plot shows the configuration rankings by the runtime of each configuration.}
    \label{fig:sup_borda_counts}
\end{figure}

To further test the robustness of the predictions, for each dimension from $16$ to $1024$, we pick 10 matrices from the test set for the classical SKQD simulations. For each of these matrices, we pick 7 configurations: the configuration with the minimum predicted error (\emph{Minimum Predicted Error}); three model-based configurations with the lowest trade-off score, computed as $\hat{\epsilon}_{prop} + 0.5\,\hat{t}$ (a heuristic metric to quantify the runtime and error tradeoff); the theoretical $\Delta t$ configuration, which combines the Hamiltonian's theoretical $\Delta t$ with the largest $m$; a random configuration, a random combination of $m$ and $\Delta t$ within the bounds of the grid; and, lastly, the configuration with the smallest state-propagation error $\epsilon_{prop}$ from the dataset construction (\emph{High Accuracy Grid}). This gives 70 simulations per dimension, which we rank per matrix to produce Borda counts~\autocite{repec:spr:sochwe:v:40:y:2013:i:2:p:353-358}. Each Hamiltonian assigns points to each of its configurations, from 7 down to 1, based on the best runtime $t$ and state-propagation error $\epsilon_{prop}$, so that the best configuration receives the most points. \autoref{fig:sup_borda_counts} shows the results as Borda counts. The \emph{Minimum Predicted Error} configuration is among the model's best configurations across all dimensions, followed closely by the \emph{Model Pick 4} configuration, while the model-picked configurations rank consistently at the top for runtime. Taking into account the trade-off between runtime and accuracy for \emph{Minimum Predicted Error} and \emph{Model Pick 4}, we pick the former to compare against the two baselines, \emph{Theoretical $\Delta t$} and \emph{Random Configuration}.

\autoref{tab:synthetic-selected-configs} reports which configurations the surrogate model actually selects. For each dimension we report, over the $10$ test matrices used in these simulations, the \emph{Minimum Predicted Error} pick, i.e.\ the $(m,\Delta t)$ with the smallest predicted state-propagation error $\hat{\epsilon}_{prop}$. The selected Krylov dimension $m$ spans the whole grid ($5$--$25$, median $15$), and the selected timestep clusters at $\Delta t\in\{0.5,1.0\}$ (median $0.5$) rather than at the smallest available value ($\Delta t=0.01$). These picks show that the surrogate learns nontrivial configuration dependence on the classical proxy; however, they are not an SKQD performance claim.

\begin{table}[htbp]
  \centering
  \caption{\textbf{Configurations selected by the surrogate on the synthetic Hamiltonians.}
  For each matrix dimension $N$ we take the $n=10$ test matrices used in the classical Krylov-Arnoldi simulations and record the \emph{Minimum Predicted Error} configuration, the $(m,\Delta t)$ pair minimizing the predicted classical state-propagation error $\hat{\epsilon}_{prop}$. We report the median and the most frequent (mode) values of the selected $m$ and $\Delta t$. The selected $\Delta t$ clusters at $0.5$--$1.0$ rather than at the grid minimum $0.01$.}
  \label{tab:synthetic-selected-configs}
  \small
  \begin{tabular}{@{}rrrrrr@{}}
    \toprule
    $N$ & $n$ & median $m$ & mode $m$ & median $\Delta t$ & mode $\Delta t$ \\
    \midrule
    16   & 10 & 15   & 15      & 0.5 & 0.5        \\
    32   & 10 & 25   & 25      & 0.5 & 0.5        \\
    64   & 10 & 12.5 & 10      & 0.5 & 0.5        \\
    128  & 10 & 17.5 & 15, 20  & 0.5 & 0.5        \\
    256  & 10 & 25   & 25      & 0.5 & 0.5        \\
    512  & 10 & 11   & 7       & 1.0 & 1.0        \\
    1024 & 10 & 10   & 10      & 0.5 & 0.05, 0.5  \\
    \midrule
    \textbf{All} & \textbf{70} & \textbf{15} & --- & \textbf{0.5} & \textbf{0.5} \\
    \bottomrule
  \end{tabular}
\end{table}

\section{Full list of hand-crafted Hamiltonian features}\label{sup:hand_crafted_matrix_features}

\begin{table}[htbp]
  \centering
  \caption{The 130 handcrafted matrix descriptors. $\mathcal{S}[\,\cdot\,]$ denotes the block of twelve summary statistics applied to the entries of its argument: mean, standard deviation, variance, minimum, maximum, skewness, excess kurtosis and the percentiles $p_{5}, p_{25}, p_{50}, p_{75}, p_{95}$. Together with the two grid parameters $(m, \Delta t)$ these form the 132 inputs of the matrix-feature surrogate. Of the 130 descriptors, 112 take distinct values on the benchmark; the remainder are constant or coincide exactly with another descriptor.}
  \label{tab:matrix-features}
  \small
  \begin{tabular}{@{}p{0.27\linewidth}p{0.58\linewidth}r@{}}
    \toprule
    Group & Descriptors & \# \\
    \midrule
    Sparsity and structure & $\mathrm{nnz}$, sparsity ratio, density, smallest nonzero magnitude, bandwidth, bandwidth/$N$ & 6 \\
    Row/column occupancy & nonzeros per row (mean, std, min, max); per column (mean, std, min, max, median, $q_{25}$, $q_{75}$, IQR, CV) & 13 \\
    Diagonal & $\mathcal{S}\!\left[\mathrm{diag}(H)\right]$ & 12 \\
    Off-diagonal (signed) & $\mathcal{S}\!\left[H_{i \neq j}\right]$, plus mean and max of $|H_{i \neq j}|$ & 14 \\
    Off-diagonal magnitude & $\mathcal{S}\!\left[\,\lvert H_{i \neq j}\rvert\,\right]$ & 12 \\
    Off-diagonal real part & $\mathcal{S}\!\left[\Re H_{i \neq j}\right]$ & 12 \\
    Off-diagonal imaginary part & $\mathcal{S}\!\left[\Im H_{i \neq j}\right]$ & 12 \\
    Off-diagonal phase & circular mean, std and variance of $\arg H_{i \neq j}$; phase uniformity; phase percentiles $p_{5},p_{25},p_{50},p_{75},p_{95}$; number of sampled phases & 10 \\
    All entries (signed) & $\mathcal{S}\!\left[H\right]$ & 12 \\
    All entries, magnitude & $\mathcal{S}\!\left[\,\lvert H\rvert\,\right]$ & 12 \\
    Norms and trace & $\lVert H\rVert_1$, $\lVert H\rVert_\infty$, $\lVert H\rVert_F$, $\mathrm{tr}\,H$, $\mathrm{tr}\,H/N$ & 5 \\
    Gershgorin bounds & lower and upper bound, width, centre & 4 \\
    Spectral (Lanczos, \texttt{eigsh}) & $\lambda_{\max}$, $\lambda_{\min}$, spectral width and radius, and the inverse spectral radius and width & 6 \\
    \midrule
    \textbf{Total} & & \textbf{130} \\
    \bottomrule
  \end{tabular}
\end{table}

\section{SqDRIFT runs on the 50 molecular Hamiltonians}\label{sup:sqdrift_runs}

\begin{figure}[htbp]
    \centering
    {\def\svgwidth{0.9\linewidth}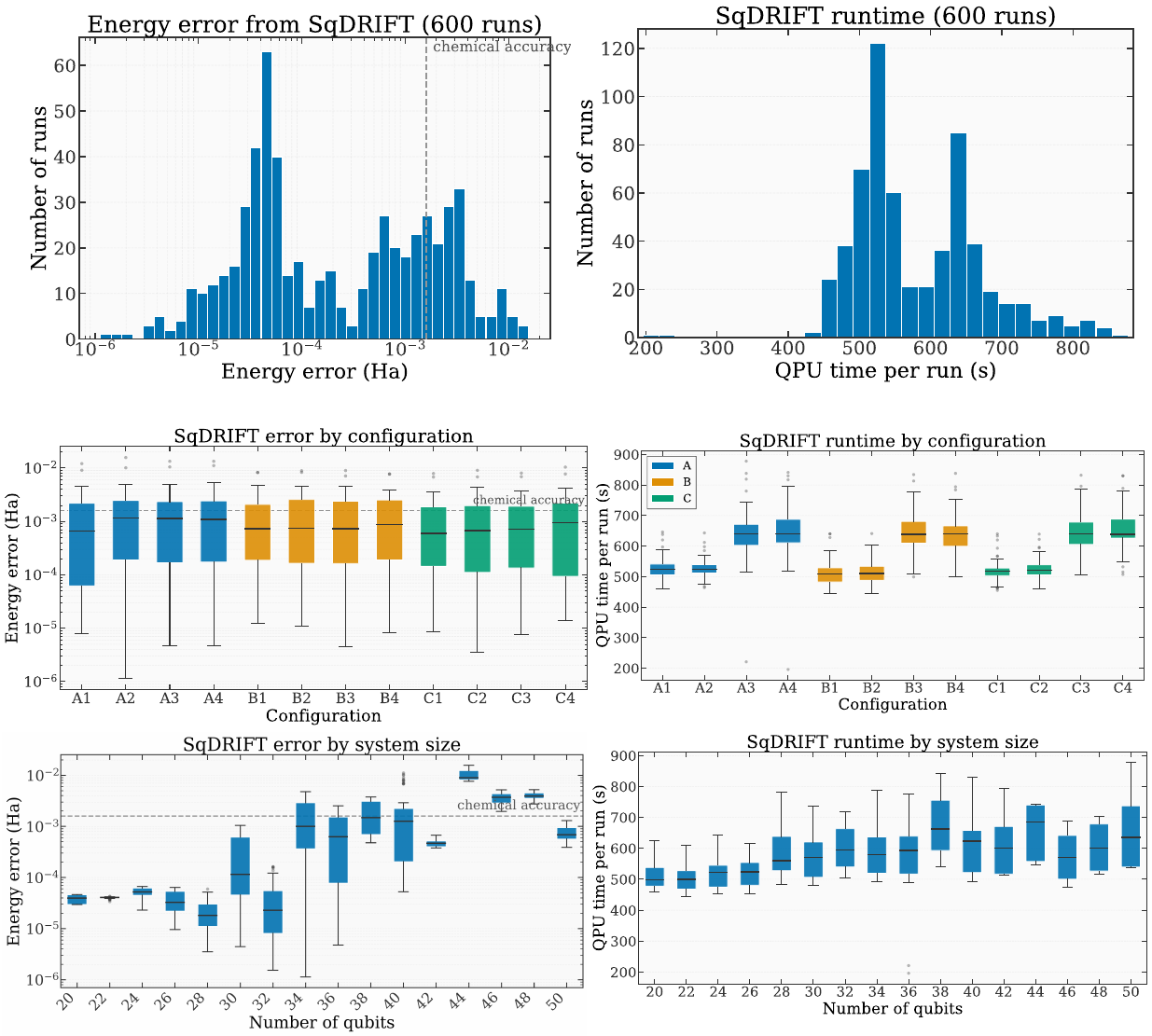}
    \caption{\textbf{The quantum hardware runs.} The first column of the figure shows the distribution of the ground state approximation errors from the SqDRIFT circuits, the error distribution per configuration and the error distribution per the number of qubits. The second column shows the runtime distributions, overall, per SqDRIFT configuration and per number of qubits. }
    \label{fig:sup_quantum_runs_data_targets}
\end{figure}

We run the experiments for the 12 SqDRIFT configurations of the 50 molecular Hamiltonians on real IBM Quantum hardware, specifically the \emph{ibm\_fez} QPU, a Heron r2 processor. The details of the configurations are given in \autoref{tab:sqdrift_configs}. In addition, \autoref{fig:sup_quantum_runs_data_targets} shows the runtime and error distributions of the targets. The SqDRIFT circuit depth for each molecule is summarized in \autoref{tab:molecules}.

\begin{table}[htbp]
\centering
\begin{tabular}{lccccc}
\toprule
\textbf{Config} & \textbf{Num.\ groups} & \textbf{Num.\ rand} & \textbf{\emph{m} ($\Delta t = 1$)} & \textbf{Num.\ samples} & \textbf{Total samples} \\
\midrule
A1 & [15] & 1000 & [1.0, 2.0, 3.0]       & 500  & 1.5M \\
A2 & [15] & 1000 & [1.0, 2.0, 3.0, 4.0]  & 375  & 1.5M \\
A3 & [20] & 1000 & [1.0, 2.0, 3.0]       & 500  & 1.5M \\
A4 & [20] & 1000 & [1.0, 2.0, 3.0, 4.0]  & 375  & 1.5M \\
B1 & [15] & 100  & [1.0, 2.0, 3.0]       & 5000 & 1.5M \\
B2 & [15] & 100  & [1.0, 2.0, 3.0, 4.0]  & 3750 & 1.5M \\
B3 & [20] & 100  & [1.0, 2.0, 3.0]       & 5000 & 1.5M \\
B4 & [20] & 100  & [1.0, 2.0, 3.0, 4.0]  & 3750 & 1.5M \\
C1 & [15] & 500  & [1.0, 2.0, 3.0]       & 1000 & 1.5M \\
C2 & [15] & 500  & [1.0, 2.0, 3.0, 4.0]  & 750  & 1.5M \\
C3 & [20] & 500  & [1.0, 2.0, 3.0]       & 1000 & 1.5M \\
C4 & [20] & 500  & [1.0, 2.0, 3.0, 4.0]  & 750  & 1.5M \\
\bottomrule
\end{tabular}
\caption{\textbf{SqDRIFT configuration grid.} The detailed combinations of parameters used in the benchmarking-molecules experiments.}
\label{tab:sqdrift_configs}
\end{table}

The ground-state energy error of the SqDRIFT runs in
\autoref{fig:sup_quantum_runs_data_targets} is bimodal: a low-error mode near $10^{-5}$\,Ha for the smallest molecules and a high-error mode for the larger ones. This reflects how much of the full configuration-interaction (FCI) space the fixed $1.5\times10^{6}$-sample budget is able to recover. \autoref{tab:subspace_coverage} reports, for each system size, the median recovered-subspace dimension (the number of merged recovered configurations passed to the SQD diagonalization step at the final iteration) relative to the FCI dimension $\binom{N_\text{orb}}{N_\uparrow}\binom{N_\text{orb}}{N_\downarrow}$ of the $(N_\text{elec}, N_\text{orb})$ active space. For the 20-qubit molecules the budget recovers $\gtrsim 96\%$ of the FCI space (e.g.\ CH$_4$, with $\binom{10}{5}^2 = 63{,}504$ determinants), so the diagonalization is effectively exact and the error is configuration-independent. The crossover where the FCI dimension exceeds the sample budget occurs around 22--24 qubits; beyond 30 qubits the recovered subspace spans less than $10\%$ of FCI, falling to $\sim10^{-5}$ for the largest molecules, where the error becomes both large and configuration-sensitive. This two-regime behaviour produces the two modes of the error distribution and explains why $\log_{10}$ of the FCI dimension is one of the strongest predictive features, as it encodes whether the sampling budget can span the relevant subspace.

\begin{table}[htbp]
  \centering
  \caption{\textbf{Recovered-subspace coverage of the FCI space, by qubit count.} For each system size we report a representative full configuration-interaction (FCI) dimension, $\binom{N_\text{orb}}{N_\uparrow}\binom{N_\text{orb}}{N_\downarrow}$, the median recovered-subspace dimension (merged recovered configurations passed to the SQD
  diagonalization step, final iteration), and their ratio. All runs use a fixed budget of $1.5\times10^{6}$ samples. Coverage falls from near-unity at 20 qubits to $\sim10^{-5}$ for the largest molecules, producing the two modes of the error distribution in
  \autoref{fig:sup_quantum_runs_data_targets}.}
  \label{tab:subspace_coverage}
  \begin{tabular}{rrrr}
    \toprule
$N_q$ & FCI dimension & Recovered (median) & Coverage \\
    \midrule
    20 & 63{,}504            & 61{,}287      & 0.97              \\
    22 & 213{,}444           & 167{,}450     & 0.78              \\
    24 & 627{,}264           & 305{,}138     & 0.49              \\
    26 & $1.7\times10^{6}$   & 472{,}286     & 0.29              \\
    28 & $4.0\times10^{6}$   & 649{,}173     & 0.16              \\
    30 & $9.0\times10^{6}$   & 665{,}982     & 0.07              \\
    32 & $1.9\times10^{7}$   & 795{,}038     & 0.04              \\
    34 & $1.5\times10^{8}$   & 1{,}054{,}910 & $7\times10^{-3}$  \\
    36 & $2.1\times10^{8}$   & 955{,}922     & $7\times10^{-3}$  \\
    38 & $1.6\times10^{9}$   & 1{,}268{,}509 & $1\times10^{-3}$  \\
    40 & $1.5\times10^{9}$   & 1{,}056{,}273 & $7\times10^{-4}$  \\
    42 & $2.9\times10^{9}$   & 1{,}072{,}253 & $4\times10^{-4}$  \\
    44 & $1.0\times10^{11}$  & 1{,}423{,}698 & $1\times10^{-5}$  \\
    46 & $6.0\times10^{10}$  & 1{,}259{,}169 & $2\times10^{-5}$  \\
    48 & $1.2\times10^{11}$  & 1{,}277{,}096 & $1\times10^{-5}$  \\
    50 & $3.1\times10^{10}$  & 1{,}101{,}623 & $4\times10^{-5}$  \\
    \bottomrule
  \end{tabular}
\end{table}

\section{Detailed results on the molecular dataset}\label{sup:ablations_molecular_hamiltonians}

\begin{figure}[htbp]
    \centering
    {\def\svgwidth{0.5\linewidth}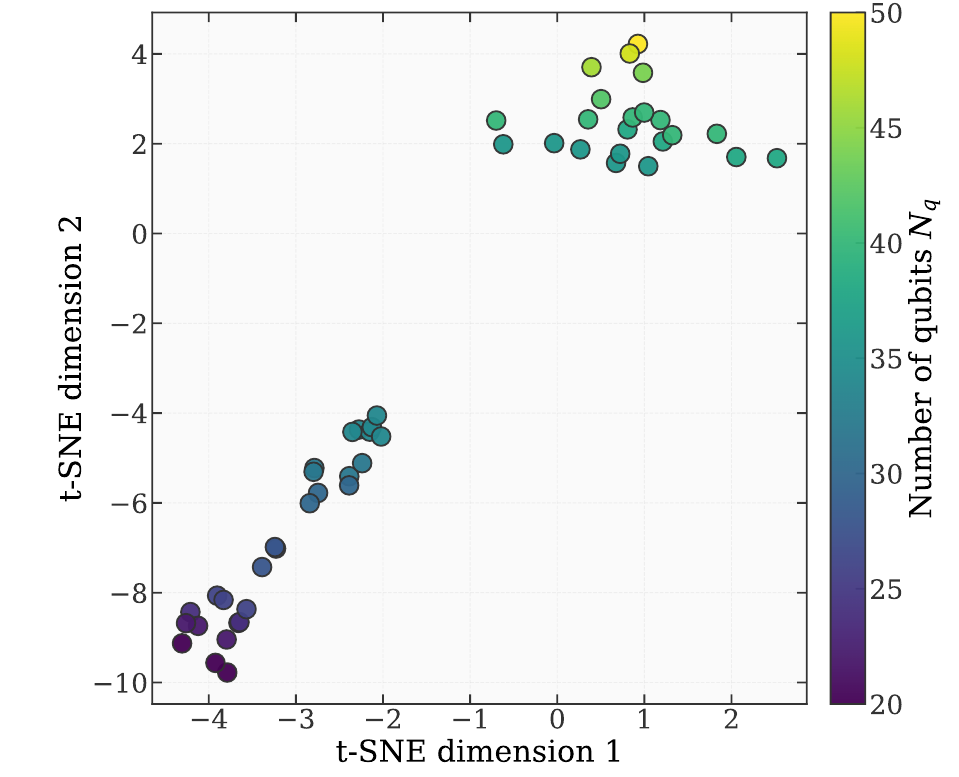}
    \caption{\textbf{t-SNE plot of the embeddings of the 50 molecular Hamiltonians.} }
    \label{fig:sup_tsne_plot_molecular_ham}
\end{figure}

\autoref{fig:sup_tsne_plot_molecular_ham} shows how the embeddings of the molecules cluster together in the 2D t-SNE reduced latent space. The qubit-count split is clear, with smaller molecules occupying the bottom-left group and larger ones occupying the top-right group.

\autoref{tab:molecules-summary} reports the model selection, 42 combinations of feature sets, estimators and baselines. The ID test set consists of 11 molecules, while the OOD test set consists of 5. The model carried forward is the one that consistently performs better than the others both on the ID and OOD test sets, on both the classification and regression tasks. The physics features combined with the 8 PCA-reduced embedding features are the only representation that performs well on both test splits: with a linear estimator they reach $\rho = 0.888$ on the ID test set and $0.784$ on the OOD test set. The physics descriptors, although they perform well on the OOD split ($\rho = 0.846$ and $0.856$ for OLS and Ridge), degrade on the ID test set ($\rho = 0.730$ and $0.723$). The embeddings alone, by contrast, perform well on the ID test set ($\rho$ up to $0.808$) but collapse on the OOD molecules, where four of the five estimators on Emb-PCA8 and three of five on the full embeddings (Emb-full) lose significance, two of them turning slightly negative. These results show that the embedding improves ID ranking (Spearman $\rho$ from 0.73 for the Physics+config OLS to 0.888 for the Physics+Emb-PCA8+config OLS) but not OOD extrapolation, while benefiting OOD in the classification task. One reason for this might be that the encoder is trained to be invariant under unitary conjugation, whereas SQD accuracy depends on the sparsity of the ground state
in the computational basis, which is exactly what conjugation changes. Isospectral Hamiltonians can have very different SQD difficulty. In addition, the context window length of the encoder allows for only 200k tokens. The number of Pauli terms for the molecular Hamiltonians ranges from around 2800 for the smaller ones to approximately 600k for the bigger OOD molecules. To circumvent this, only the 200k Pauli terms $P_j$ with the highest coefficients $c_j$ were used for the embeddings.

On every physics-bearing representation, the linear estimators dominate the tree ensembles on the OOD test set: for Physics+Emb-PCA8 the gap is $0.784$ against $0.191$–$0.577$, even though the ensembles match them on the ID test set ($\rho \ge 0.855$). This mirrors the behaviour on the synthetic benchmark: a forest cannot extrapolate beyond the range of targets it was trained on, whereas a linear model in log space can. The same model also answers the classification framing. Fitted on Physics+Emb-PCA8+config and read through its normal-CDF head, using the in-sample residual variance of the OLS fit in natural-log space as the variance, it attains the highest balanced accuracy of that feature set on both splits, $0.618$ in distribution and $1.000$ on the five OOD molecules, against $0.500$–$0.590$ for the five estimators trained directly on the binary label. We note that the feature-free comparison in \autoref{tab:molecules-summary} makes the dominant signal explicit: a classifier using only physics descriptors and no configuration features already reaches $0.970$ balanced accuracy on the ID split, confirming that the chemical-accuracy decision is governed primarily by the molecule (size, electron count, Hilbert-space dimension) rather than by the configuration. The configuration-aware model is therefore best read as an \emph{a priori feasibility and resource estimator}—predicting whether a given molecule is tractable on this hardware and at what runtime—rather than as a fine-grained per-configuration selector for the molecular dataset. Overall, an OLS fit on $\log|\epsilon|$ serves both the ranking of configurations and the chemical-accuracy decision. The detailed results of the selected model appear in \autoref{sec:results}.

\begin{table}[htbp]
  \centering
  \caption{\textbf{Molecular Hamiltonians model selection.} The classification task learns and predicts whether a configuration yields an error below or above chemical accuracy at $1.6\times10^{-3}$ Ha and is evaluated using \emph{Balanced accuracy}. The regression task learns and predicts the QPU runtime $t$ and the error $\epsilon$, and is evaluated using \emph{Spearman }$\rho$, the rank correlation between true and predicted values. Feature sets all include the six circuit configuration parameters (config) unless otherwise noted; ``physics'' features include molecule data such as the no. of electrons, the no. of orbitals and the $\log$ of the full Hilbert space dimension. OLS pairs the regressor with its normal-CDF classification head, so the two columns of that row share one underlying model. Bold marks the models carried forward, $^\dagger$ a $\rho$ not significant at $\alpha = 0.05$. The last two rows are the feature-free classification baselines; a balanced accuracy of exactly $0.500$ means the classifier assigns every sample to one class.}
  \label{tab:molecules-summary}
  \small
  \setlength{\tabcolsep}{5pt}
  \begin{tabular}{llrrrr}
    \toprule
    & & \multicolumn{2}{c}{Balanced accuracy (chance $0.500$)} & \multicolumn{2}{c}{Spearman $\rho$ (null $0$)} \\
    \cmidrule(lr){3-4}\cmidrule(lr){5-6}
    Feature set & Estimator & ID test & OOD test & ID test & OOD test \\
    \midrule
    Physics + config & OLS & 0.727 & 0.521 & 0.730 & 0.846 \\
     & Ridge & N/A & N/A & 0.723 & 0.856 \\
     & RF & 0.617 & 0.542 & 0.642 & 0.643 \\
     & XGBoost & 0.650 & 0.528 & 0.626 & 0.774 \\
     & LightGBM & 0.655 & 0.500 & 0.608 & 0.699 \\
     & LogReg & 0.684 & 0.667 & N/A & N/A \\
     & SVM & 0.604 & 0.660 & N/A & N/A \\
    \midrule
    Emb-PCA8 + config & OLS & 0.548 & 0.500 & 0.654 & -0.099$^\dagger$ \\
     & Ridge & N/A & N/A & 0.663 & -0.030$^\dagger$ \\
     & RF & 0.500 & 0.500 & 0.704 & 0.173$^\dagger$ \\
     & XGBoost & 0.448 & 0.500 & 0.762 & 0.066$^\dagger$ \\
     & LightGBM & 0.564 & 0.500 & 0.808 & 0.339 \\
     & LogReg & 0.530 & 0.500 & N/A & N/A \\
     & SVM & 0.396 & 0.583 & N/A & N/A \\
    \midrule
    Emb-full + config & OLS & 0.309 & 0.500 & -0.246 & 0.733 \\
     & Ridge & N/A & N/A & -0.257 & 0.733 \\
     & RF & 0.564 & 0.500 & 0.661 & 0.234$^\dagger$ \\
     & XGBoost & 0.534 & 0.500 & 0.625 & 0.216$^\dagger$ \\
     & LightGBM & 0.512 & 0.493 & 0.653 & 0.207$^\dagger$ \\
     & LogReg & 0.564 & 0.500 & N/A & N/A \\
     & SVM & 0.396 & 0.417 & N/A & N/A \\
    \midrule
    Physics + Emb-PCA8 + config & \textbf{OLS} & \textbf{0.618} & \textbf{1.000} & \textbf{0.888} & \textbf{0.784} \\
     & Ridge & N/A & N/A & 0.888 & 0.779 \\
     & RF & 0.617 & 0.500 & 0.884 & 0.476 \\
     & XGBoost & 0.599 & 0.576 & 0.855 & 0.191$^\dagger$ \\
     & LightGBM & 0.573 & 0.590 & 0.883 & 0.577 \\
     & LogReg & 0.564 & 0.500 & N/A & N/A \\
     & SVM & 0.448 & 0.583 & N/A & N/A \\
    \midrule
    \multicolumn{6}{l}{\emph{Classification only}} \\[2pt]
    Config & OLS & 0.559 & 0.500 & N/A & N/A \\
     & RF & 0.559 & 0.500 & N/A & N/A \\
     & XGBoost & 0.559 & 0.500 & N/A & N/A \\
     & LightGBM & 0.559 & 0.500 & N/A & N/A \\
     & LogReg & 0.525 & 0.500 & N/A & N/A \\
     & SVM & 0.492 & 0.500 & N/A & N/A \\
    Physics (no config) & OLS & 0.853 & 0.500 & N/A & N/A \\
     & RF & 0.970 & 0.500 & N/A & N/A \\
     & XGBoost & 0.970 & 0.500 & N/A & N/A \\
     & LightGBM & 0.970 & 0.500 & N/A & N/A \\
     & LogReg & 0.853 & 0.750 & N/A & N/A \\
     & SVM & 0.801 & 0.500 & N/A & N/A \\
    \midrule
     & Majority & 0.500 & 0.500 & N/A & N/A \\
     & Prevalence & 0.500 & 0.500 & N/A & N/A \\
    \bottomrule
  \end{tabular}
\end{table}

\begin{figure}[htbp]
    {\def\svgwidth{\linewidth}
\begingroup%
  \makeatletter%
  \providecommand\color[2][]{%
    \errmessage{(Inkscape) Color is used for the text in Inkscape, but the package 'color.sty' is not loaded}%
    \renewcommand\color[2][]{}%
  }%
  \providecommand\transparent[1]{%
    \errmessage{(Inkscape) Transparency is used (non-zero) for the text in Inkscape, but the package 'transparent.sty' is not loaded}%
    \renewcommand\transparent[1]{}%
  }%
  \providecommand\rotatebox[2]{#2}%
  \newcommand*\fsize{\dimexpr\f@size pt\relax}%
  \newcommand*\lineheight[1]{\fontsize{\fsize}{#1\fsize}\selectfont}%
  \ifx\svgwidth\undefined%
    \setlength{\unitlength}{588.18547286bp}%
    \ifx\svgscale\undefined%
      \relax%
    \else%
      \setlength{\unitlength}{\unitlength * \real{\svgscale}}%
    \fi%
  \else%
    \setlength{\unitlength}{\svgwidth}%
  \fi%
  \global\let\svgwidth\undefined%
  \global\let\svgscale\undefined%
  \makeatother%
  \begin{picture}(1,0.29564558)%
    \lineheight{1}%
    \setlength\tabcolsep{0pt}%
    \put(0.00047244,0.26574002){\color[rgb]{0,0.03137255,0.21568627}\makebox(0,0)[lt]{\lineheight{0}\smash{\begin{tabular}[t]{l}\textbf{A}\end{tabular}}}}%
    \put(0,0){\includegraphics[width=\unitlength,page=1]{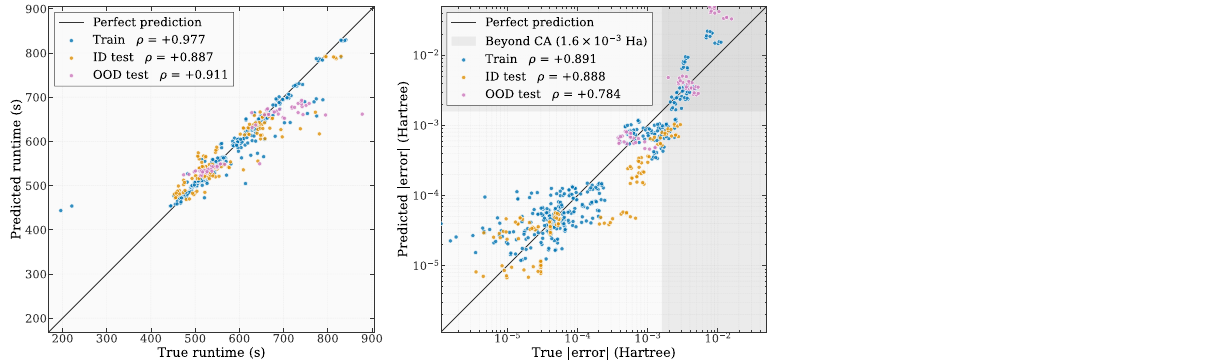}}%
    \put(0.63192075,0.26611528){\color[rgb]{0,0.03137255,0.21568627}\makebox(0,0)[lt]{\lineheight{0}\smash{\begin{tabular}[t]{l}\textbf{B}\end{tabular}}}}%
    \put(0,0){\includegraphics[width=\unitlength,page=2]{real_ham_results_panel_svg-tex.pdf}}%
  \end{picture}%
\endgroup%
}
    \caption{\textbf{Summary results on the molecular Hamiltonians' dataset}. Panel A shows the predicted vs true runtime and absolute values of errors for the three sets: train, in-distribution test and out-of-distribution test set. The shaded area in the errors plot highlights the samples whose true and predicted errors fall beyond chemical accuracy (CA=$1.6 \times 10^{-3}$). Panel B shows the Spearman correlation coefficient and the median absolute percentage error (med. APE) between the true and predicted runtimes, and the Spearman correlation coefficient between the true and predicted absolute errors on the in-distribution and out-of-distribution test sets, together with the balanced accuracy for classifying whether a configuration's prediction falls below or above chemical accuracy.}
    \label{fig:chemical_ham_full_results_best_model}
\end{figure}

\autoref{fig:chemical_ham_full_results_best_model} shows the results, on all three sets, of the random forest model for the runtime prediction and the Ordinary Least Squares model for the ground state energy approximation error.

\end{appendices}

\end{document}